\documentclass[preprint,12pt]{elsarticle}

 \biboptions{comma,square}

\usepackage[T1]{fontenc}
\usepackage[latin9]{inputenc}
\usepackage{verbatim}
\usepackage{amstext}
\usepackage{color}
\usepackage{graphicx,psfrag}
 \usepackage{amssymb}

\def\eg{\emph{e.g. }}
\def\ie{\emph{i.e. }}
\def\aka{\emph{a.k.a. }}
\def\beq{\begin{equation}}
\def\eeq{\end{equation}}
\def\beqa{\begin{eqnarray}}
\def\eeqa{\end{eqnarray}}
\def\beg{\begin{lyxgreyedout}}
\def\eeg{\end{lyxgreyedout}}
\def\nn{\nonumber}
\def\eps{{\epsilon}}

\begin{document}

\begin{frontmatter}

\title{Classical geometry to quantum behavior correspondence in a \textit{Virtual} Extra Dimension}

  \author{Donatello Dolce}
  \ead{ddolce@unimleb.edu.au}

\date{\today}
\address{Centre of Excellence for Particle Physics (CoEPP),\\  The University of Melbourne,\\   Parkville 3010 VIC, \\ Australia.
}

\begin{abstract}

In the Lorentz invariant formalism of compact space-time dimensions the assumption of periodic boundary conditions represents a consistent semi-classical quantization condition for relativistic fields. 
In \cite{Dolce:2009ce} we have shown, for instance, that the ordinary Feynman path integral is obtained from the interference between the classical paths with different winding numbers associated with the cyclic dynamics of the field solutions. 
By means of the boundary conditions, the kinematics information of interactions can be encoded on the relativistic geometrodynamics of the boundary \cite{Dolce:tune}.
Furthermore, such a purely four-dimensional theory is manifestly dual to an extra-dimensional field theory.
The resulting  correspondence between extra-dimensional geo\-metrodynamics and ordinary quantum behavior can be interpreted in terms of  AdS/CFT correspondence.  
By applying this approach to a simple  Quark-Gluon-Plasma  freeze-out model we obtain fundamental analogies with basic aspects of AdS/QCD phenomenology. 
\end{abstract}

\end{frontmatter}
\begin{keyword}
{Compact Dimensions, Kaluza-Klein, Semi-classical methods, Relativistic Geometrodynamics, AdS/CFT,  Quark-Gluon-Plasma}
\end{keyword}

\addcontentsline{toc}{section}{Introduction}

\section*{Introduction\label{sec:0}}
Originally proposed by T. Kaluza \cite{Kaluza:1921tu} and O. Klein
\cite{Klein:1926tv}, field theory in compact eXtra Dimension (XD)  represents one of the most investigated
candidates for new physics beyond the Standard Model (SM), providing an
elegant explanation for the hierarchy problem \cite{ArkaniHamed:1998rs,Randall:1999vf,Schmaltz:2002wx,Csaki:2003dt}.

Nevertheless, Kaluza, following Nordstr\"om's idea \cite{Nordstrom:1914}, originally introduced the XD formalism as a ``mathematical trick'' to unify  gravity  and electromagnetism \cite{Kaluza:1921tu}. This mathematical property of XD theories has inspired the purely 4D, semi-classical, geometrodynamical formulation of scalar QED recently published in \cite{Dolce:tune}, in which gauge symmetries are related to corresponding space-time symmetries.

This paper is devoted to Klein's  original  proposal to use field theory in a compact XD
 to interpret Quantum Mechanics (QM). In
his famous paper \emph{Quantentheorie und f\"unfdimensionale Relativit\"atstheorie}
\cite{Klein:1926tv} Klein noted that Periodic Boundary Conditions
(PBCs) at the ends of a compact XD yield an analogy to the Bohr-Sommerfeld
quantization condition ---  the cyclic XD was introduced to interpret the quantization of the electric charge. 
 As we will see in detail, the formalism of XD theory shares interesting analogies with the quantum formalism. 
For instance, in XD theories the KK mass spectrum is fixed by consistent
Boundary Conditions (BCs) similarly to the energy spectrum in semi-classical theories, whereas the evolution along
the XD is described by bulk Equations of Motion (EoMs) playing the role
of the Schr\"odinger equation of QM, \cite{Dvali:2001gm,Carena:2002me}. 
 In general the essential requirement for the BCs is that they must minimize the action at the boundaries, fulfilling the classical variational principle \cite{Henneaux:1998ch,Csaki:2003dt}.
It is  well known from XD field theory or string theory that the variational
principle allows PBCs (or anti-PBCs) in relativistic bosonic actions  --- or in an Orbifold description   combinations of BCs of Neumann (N-BCs) and Dirichlet (D-BCs)  type. These BCs have the same formal validity of the usual Synchronous (\aka Stationary) BCs (SBCs) of ordinary field theory, in which the fields have fixed values at
the boundaries. 

The dimensional (de)construction mechanism
 \cite{Arkani-Hamed:2001ca,Son:2003et}
has shown another remarkable application of XD theories, that is to say a dualism between the classical configuration of XD fields, in particular in warped XD (AdS),  and the quantum behavior of Strongly Coupled Field Theories (SCFTs), such as QCD  \cite{Pomarol:2000hp,ArkaniHamed:2000ds}.
This so-called AdS/QCD correspondence is a phenomenological application
of the more fundamental (and not yet proven) AdS/CFT correspondence \cite{Maldacena:1997re}, whose validity seems to be more general than its original formulation, as confirmed by many bottom-up investigations. 
According to the  words used by  Witten in the abstract of his famous paper \cite{Witten:1998zw}, in AdS/CFT
 ``\emph{quantum phenomena [...] are encoded in classical geometry}''. 
This aspect can be regarded as a ``classical to quantum'' correspondence
which, however, does not involve any explicit quantization condition.

In this paper we will seek the origin of this kind of  ``classical to quantum'' correspondence of XD theories  in term of the formalism of bosonic field theory in compact space-time dimensions (compact 4D). 
  The foundational aspects of this novel theory have been defined in recent papers \cite{Dolce:2009ce,Dolce:tune}, see also \cite{Dolce:2009cev4,Dolce:QTRF5,Dolce:2010ij,Dolce:Dice,Dolce:FQXi, Dolce:Cyclic}, and will be summarized in sec.(\ref{Covariant:Notation}-\ref{sec:Towards-Interactions}). 
The basic assumption is that \emph{every (free) elementary relativistic particle has associated an intrinsic (persistent) periodicity}. This hypothesis was originally introduced by de Broglie in 1923, in terms of  intrinsic ``\emph{periodic phenomenon allied with every parcel of energy}'' whose ``\emph{character is yet to be determined}'' \cite {Broglie:1924}, \ie a so-called ``internal clock''  associated to every elementary particle.    
 In other words we will enforce the undulatory nature of elementary particle (wave-particle duality) at the base of the modern description of QM, implicitly tested  in 80 years of QFT and indirectly observed in a recent experiment \cite{2008FoPh...38..659C}. As is well known, through the Planck constant, the space-time periodicity of a de Broglie ``periodic phenomenon'' (whose minimal topology is $\mathbb S^{1}$)  describes the energy-momentum of a corresponding particle. In order to see the is full consistency with relativity of the assumption of intrinsic periodicity,  it is therefore sufficient to note that  the retarded and local variations of energy-momentum characterizing relativistic interaction and causality can be equivalently described by corresponding  retarded and local modulations of  space-time periodicity.    
 
  Essentially, by considering that in classical-relativistic mechanics the kinematics of a particle is described by its 4-momentum, that in ordinary undulatory mechanics the energy-momentum  is encoded in temporal and spatial periodicity of a corresponding intrinsically ``periodic phenomenon'' (wave-particle duality), and that in the atomistic description of modern physics elementary particles are the basic constituents of every physical system, we can figure out the possibility of a consistent description of physics in terms of intrinsically cyclic elementary systems (with locally modulated periodicity in the interacting case). In addition to this, we must consider that massless fields (such as the mediators of the EM and gravitational interactions) can have infinite periodicity, providing the long  space-time reference scales for a relational description of the elementary particles.  
Field theory in compact 4D,   is a natural realization of the de Broglie  ``periodic phenomenon'', \cite{Dolce:2009ce,Dolce:tune}. In fact, it is essentially a relativistic bosonic theory in which, in agreement with the variational principle,  PBCs are imposed at the ends of compact 4D. As a consequence, the resulting field solution is constrained to have intrinsic space-time periodicity. In turn, the de Broglie spacial and temporal periodicities are encoded in the spatial and temporal compactification lengths respectively, \ie on the boundary of the theory through the PBCs. Hence, according to undulatory mechanics, this will allow us to describe relativistic interactions in terms of corresponding  retarded and local deformations of  the compact 4D of the theory \cite{Dolce:tune}.

The assumption of intrinsic periodicity represents a semi-classical quantization condition for relativistic particles. Field theory in compact 4D can be regarded as the full relativistic generalization  of the quantization of a ``particle in a  box'' or of a ``vibrating string'', or more in general of sound theory. Moreover it has fundamental analogies with the Matsubara theory   \cite{Matsubara:1955ws}.
 Our periodic field, owing to PBCs, is nothing but a relativistic string vibrating in a compact 4D. 
 Through discrete Fourier transform and the Planck constant,  space-time periodicity \emph{directly} implies a discretized (quantized) energy-momentum spectrum. Transformations of reference frame imply relativistic modulation of space-time periodicity (\eg as in the relativistic Doppler effect).   In this way we find out that a periodic field solution has the same energy-momentum spectrum as an ordinary second quantized bosonic field (after normal ordering).

As proven in  \cite{Dolce:2009ce,Dolce:tune,Dolce:2009cev4}, without any explicit quantization condition except intrinsic periodicity, field theory in compact 4D reproduces formally  the fundamental aspects of QM such as Hilbert space, Schr\"odinger equation, commutation relations, Heisenberg relation and so on. 
In particular in sec.(\ref{sec:Correspondence-with-QFT}) we will show that  the Feynman Path Integral (FPI) formulation arises in a semi-classical and intuitive way in terms of interference between classical paths of different winding numbers associated with the underlying cyclic geometry $\mathbb S^{1}$ of the ``periodic phenomenon'', \cite{Dolce:2009ce,Dolce:2009cev4}.  Remarkably, in  \cite{Dolce:tune} we have extended such a semi-classical formulation of quantum-relativistic mechanics to scalar QED.

 In this paper, we will explore another interesting property of field theory in compact 4D. This is a remarkable mathematical analogy of our theory with a KK theory, as already pointed out in \cite{Dolce:2009ce}. In sec.(\ref{sec:Virtual-Extra-Dimension}) we will see, for instance, that the parameterization of the  intrinsic periodicity of an ``internal clock'' enters into the equations in perfect analogy with a cyclic XD of corresponding periodicity. In other words, at a mathematical level, the intrinsic periodicity of a de Broglie ``periodic phenomenon'', whose topology $\mathbb S^{1}$ is the same topology of a KK theory,  appears to be parameterized  by cyclic XD, though the theory is purely 4D. To show this we will start with an ordinary KK theory and use the cyclic XD to parameterize the cyclic world-line of a ``periodic phenomenon''.  To address this parameterization we  say that the  XD is \emph{Virtual} (VXD). 
 In particular, as already described in  \cite{Dolce:2009ce}, we will find out that under  this  identification, the KK field theory turns out to be equivalent to  field theory in compact 4D, \ie the two theories are \emph{dual}.    
Also, with the term \emph{virtual} we want to emphasize the fact, owing to this duality, that field theory in compact 4D inherits important mathematical properties from the XD theories without involving any ``real'' XD. 
For instance, in the ordinary XD the KK modes  describe \emph{independent} particles. In the formalism of compact 4D the KK mode will correspond to the collective energy eigenmodes  of the same 4-periodic field (similarly to the thermal modes of the Matsubara theory). According to the correspondence  to QM, this energy eigenmodes can be interpreted as quantum excitations or, loosely speaking, as \emph{virtual} particles, so that  we will for instance say that, in field theory in compact 4D, the KK modes are \emph{virtual}.  We will note that a similar description of the KK modes is implicitly obtained through the Holography prescription.

In sec.(\ref{sec:Towards-Interactions})  we will generalize the theory to curved 4D in order to achieve a geometrodynamical description of interaction, similar to GR, see \cite{Dolce:tune}. The local variations of 4-momentum of a given interaction scheme can be equivalently described as  
the corresponding local modulation of space-time periodicity, according to undulatory mechanics.   As already noticed in \cite{Dolce:tune}, through the PBCs of  our theory, this modulation can be equivalently formalized as corresponding local stretching of the compact 4D, that is to say in  corresponding local deformations of the underlying space-time metric. We will see that this formulation of interaction 
mimics the ordinary geometrodynamic description of linearized gravity. For instance, it is well known that in General Relativity (GR) the deformations of space-time associated to gravitational interaction encode the modulations of space-time periodicity of reference lengths and clocks, \cite{Ohanian:1995uu}.  Moreover, \cite{Dolce:tune}, in this formalism the kinematical information of the particle is encoded in the geometrodynamics of the boundary in the manner of the holographic principle \cite{'tHooft:1993gx,Susskind:1994vu}.

Combining the correspondence of the theory in compact 4D with ordinary QFT, described in \cite{Dolce:2009ce,Dolce:tune} and summarized in sec.(\ref{sec:Correspondence-with-QFT}), and the dualism to  XD theory, described  in \cite{Dolce:2009ce} and in sec.(\ref{sec:Virtual-Extra-Dimension}), together with  the geometrodynamical formulation of interactions, described in  \cite{Dolce:tune} and summarized in sec.(\ref{sec:Towards-Interactions}), we  will infer  that \emph{the classical configuration of  fields in a deformed VXD background  encodes  the quantum behavior of the corresponding
interaction scheme}. In sec.(\ref{sec:AdSCFT:interpr}) we will apply this idea to a simple Bjorken Hydrodynamical Model (BHM) for Quark-Gluon-Plasma (QGP) exponential freeze-out \cite{Magas:2003yp}, and---considering also the analogies of QCD with a thermodynamic system \cite{Satz:2008kb}---we will find a parallelism with basic aspects of AdS/QCD phenomenology. In particular, during the exponential free-out the 4-momentum of the QGP fields decays exponentially---in terms of thermal QCD this corresponds to Newton's law of cooling. Thus the de Broglie ``internal clocks'' space-time periodicity during the freeze-out has an exponential dilatation (modulation). This in turn can be equivalently encoded in a \emph{virtual} AdS metric. By solving the classical propagation of the field in this VXD geometry we find that: gauge coupling has a logarithmic running typical of the asymptotic freedom; the energy excitations (\ie \emph{virtual} KK modes) of the de Broglie ``periodic phenomena'' can be interpreted as hadrons; and the time periodicity turns out to be the conformal parameter of AdS/CFT, so that it naturally parameterizes the inverse of the energy of the QGP. 
 We  conclude that the classical geometry to quantum behavior correspondence of field theory in compact 4D has important justifications in basic aspects of  AdS/CFT phenomenology,  providing at the same time an unconventional interpretation of Maldacena's conjecture in terms of the wave-particle duality and space-time geometrodynamics. 
 
Every section of this paper is concluded by comments and outlooks, providing physical interpretations of the formal results. 

\section{Intrinsic periodicity}\label{Covariant:Notation}

In this section we define the basic aspects of field theory in
compact 4D, see \cite{Dolce:2009ce,Dolce:tune}. 
The results are presented
for scalar bosonic fields, so that they can be easily generalized to vector bosons --- the generalization to fermions will be given in future papers.   

The undulatory nature of elementary particles was introduced by de Broglie in 1923 in these terms: ``\emph{we proceed in this work from the assumption of the existence of a
  certain periodic phenomenon of a yet to be determined character, which is to
  be attributed to each and every isolated energy parcel [elementary
  particle]}'', \cite{Broglie:1924}. This ``\emph{yet to be determined}''  de Broglie ``periodic phenomenon'' has been tested by 80 years of successes of QFT  and observed indirectly in a recent experiment \cite{1996FoPhL,2008FoPh...38..659C}.
According to de Broglie,  a particle with mass $\bar M$  can be characterized by a proper-time periodicity $T_{\tau}$ defined by the relation $ T_\tau \equiv h / \bar M c^2$, \ie by a ``periodic phenomenon'' with minimal topology $\mathbb S^1$.   
 Depending on the reference frame, such a proper-time periodicity $T_\tau$  induces corresponding spatial and temporal periodicities  $\vec \lambda_{x}$ and  $T_{t}$ respectively, as can be easily checked through Lorentz transformation, 
\begin{equation}c T_\tau= c \gamma T_{t} - \gamma \vec \beta \cdot \vec \lambda_{x}\,.\label{Lorentz:prop:time} \end{equation}  
It is convenient to write the resulting de Broglie space-time periodicity  in a covariant notation  
\begin{equation}
T_{\mu}=\{T_{t},-\vec \lambda_{x}\mathbf{\hat{n}}/c\}\,.
\end{equation}
As is well known from undulatory mechanics, $T^{\mu}$ can be used to describe the 4-momentum $\bar p_{\mu}=\{\bar E /c,-\mathbf{\bar{p}}\}$ of an elementary particle with mass $\bar M$. In fact, in the new reference frame denoted by $\mathbf{\bar{p}}$, the resulting energy and momentum are  $\bar E(\mathbf{\bar{p}}) = \gamma \bar M c^{2}$ and $\mathbf{\bar{p}} = \gamma \vec {\beta} \bar M c$ respectively, so that the de Broglie relation (\aka de Broglie phase harmony) can be in general rewritten as
\begin{equation}
   T_\tau \bar M c^2 \equiv h ~~~~~~~ \rightarrow ~~~~~~~  T^{\mu} {\bar{p}}_{\mu} c \equiv h\,. \label{eq:4D:debroglie:period}
\end{equation}
In particular we have the familiar relation $\bar p_{\mu}=\hbar \bar \omega_{\mu} / c $ where  $\bar{\omega}_{\mu}$ is the fundamental 4-angular-frequency associated with $T^\mu$.

The periodicity  $T_{\tau}$ of the proper-time $\tau$ can be equivalently parameterized  by a world-line parameter $s=c \tau$ of corresponding periodicity
\begin{equation}
\lambda_{s}= c T_{\tau} =\frac{h}{\bar{M}c}\label{Compton:mass:period}\,.
\end{equation}
We note that $\lambda_{s}$ is nothing other than the Compton wavelength of the corresponding particle of mass $\bar M$. Therefore we find out that massive particles  (except neutrinos) have typically extremely fast periodic dynamics if compared with the characteristic periodicity of the Cs atom  $T_{Cs} \sim 10^{-10} s$, or with  the present  experimental resolution in time $\Delta T_{exp} \sim 10^{-17} s$. For instance a hypothetical light boson with the mass of an electron has intrinsic proper-time periodicity  $T_{\tau} \sim 10^{-20} s$, whereas an hypothetical TeV boson has intrinsic proper-time periodicity  $T_{\tau} \sim 10^{-27} s$  --- a possible generalization of the de Broglie ``periodic phenomenon''   to fermionic fields is naturally represented by  the \emph{Zitterbewegung} model ideated by Schr\"odinger. In sec.(\ref{sec:Correspondence-with-QFT}) we will show that these extremely small time scales are relevant for interpreting the quantum behavior of elementary particles  \cite{Dolce:2009ce,'tHooft:2001ar, Elze:2003tb}.

\subsection{Compact space-time}\label{Sec:Compact:spacetime}

We want to formalize the de Broglie assumption of a ``periodic phenomenon'' by describing an isolated elementary bosonic particle of classical 4-momentum  $\bar{p}_{\mu}$ as a bosonic field solution $\Phi(x)$ of persistent de Broglie space-time periodicity $T^{\mu}$ imposed as a constraint, according to (\ref{eq:4D:debroglie:period}). That is, suppressing the Lorentz index, such a de Broglie ``periodic phenomenon'' is represented as the bosonic solution  
\begin{equation}
\Phi(x)=\Phi(x+Tc)\,.\label{eq:cond:4D:period}
\end{equation}

 Furthermore we note that every isolated elementary particle, \ie constant 4-momentum, has associated persistent periodicity, \ie constant $T^\mu$. Therefore it can be regarded as a reference clock, \aka the ``de Broglie internal clock''. In fact, according to Einstein, " \emph{by a [relativistic] clock we understand anything characterized by a
  phenomenon passing periodically through identical phases so that we must
  assume, by the principle of sufficient reason, that all that happens in a
  given period is identical with all that happens in an arbitrary period}", \cite{Einstein:1910}. 
 That is, as in Einstein's relativistic clock, in a  de Broglie "periodic phenomenon" the whole physical information is contained in a single space-time period.     
Hence we can represent a de Broglie ``periodic phenomenon'' as the bosonic field (\ref{eq:cond:4D:period}) of intrinsic periodicity $T^{\mu}$, \ie as the  solution of the following bosonic action in compact 4D and PBCs 
  \begin{equation}
  {\mathcal{S}}^{\lambda_{s}} = \oint^{T^\mu} d^4 x {\mathcal{L}}(\partial_\mu \Phi(x),\Phi(x))\label{generic:actin:comp4D}\,.
\end{equation}
In this notation the PBCs are represented by the circle in the integral symbol $\oint$. As is well known from XD theory, PBCs (\ref{eq:cond:4D:period})---or  combinations of N BCs or D BCs---are fully consistent with the above bosonic action in compact dimensions, since they minimize the action at the boundary fulfilling the variational principle. Note that in ordinary field theory this requirement is the motivation for the assumption of the stationary field (fixed value) at the
ends of a time interval, \emph{i.e.} SBCs. 
Both SBCs and PBCs are equivalently allowed by the variational
principle. In particular they  both preserve the relativistic symmetries of a bosonic theory, see par.(\ref{relativistic:geometrodynamics}). This is the fundamental reason why  field theory in compact 4D is a fully consistent relativistic theory.

Here it is sufficient to note that, by using a global Lorentz transformation \begin{equation}x_\mu \rightarrow x'_\mu = \Lambda_\mu^\nu x_\nu \,,\label{Lorentz:transf}\end{equation} as a linear transformation of variables in the action (\ref{generic:actin:comp4D}), we get the transformed action
 \begin{equation}
  {\mathcal{S}}^{\lambda_{s}} = \oint^{\Lambda^\mu_\nu T^\nu} d^4 x' {\mathcal{L}}( \partial'_\mu \Phi'(x'),\Phi'(x'))\,.\label{generic:actin:comp4D:transf}
\end{equation}
Besides the transformation of the integrand, a related transformation of the boundary of the theory must be considered. 
Thus, \cite{Birrell:1982ix}, the compactification length $T^\mu$ of the action  transforms  in  contravariant  way with respect to (\ref{Lorentz:transf}),   
\begin{equation} 
T^\mu \rightarrow {T'}^\mu = \Lambda^\mu_\nu T^\nu\,. \label{Lorentz:transf:T}
\end{equation}  
This also means that the field solution $\Phi'(x')$ minimizing the transformed action (\ref{generic:actin:comp4D:transf})  has intrinsic persistent periodicity  $T'^\mu$. That is, in the new reference frame  the resulting space-time periodicity $T'^\mu$ of the field  $\Phi'(x')$ solution  of (\ref{generic:actin:comp4D:transf}) is transformed in a contravariant way.  Indeed, according to  the de Broglie phase harmony (\ref{eq:4D:debroglie:period}), it actually describes the transformed 4-momentum of the particle of mass $\bar M$ in the transformation of reference frame (\ref{Lorentz:transf}). In fact,    
\begin{equation}
\bar p_\mu \rightarrow  {{\bar p}'}_\mu = \Lambda_\mu^\nu \bar p_\nu\,. \label{Lorentz:transf:X}
\end{equation} 

In this formalism the  covariance of the theory in  compact 4D  is manifest \cite{Dolce:2009ce}. 
As a double check  we can also note that, according to (\ref{eq:4D:debroglie:period}), the phase of a field is invariant under translations of de Broglie 4-periods $T^{\mu}$ and it must also be a scalar under Lorentz transformations  \cite{Dolce:2009ce,Broglie:1924,1996FoPhL}.

The consistency of this covariant formalism  can also be  seen by noticing that, as well known from undulatory mechanics, see (\ref{eq:4D:debroglie:period}),  
the local and retarded variations of energy-momentum    $\bar p_{\mu}$ occurring during interaction can be equivalently interpreted as  local and retarded modulations of de Broglie space-time periodicity $T^{\mu}$.  Indeed in par.(\ref{sec:Towards-Interactions}) we will describe interactions in terms of modulation of periodicity. 
More precisely, it is possible to show that the de Broglie space-time periodicity $T^{\mu}$  transforms as is a covariant ``tangent''\footnote{It must be noticed that the local (``instantaneous'' in the modulated wave terminology) 4-periodicity of a field  is in general different from the compactification 4-length of the local action. Under a generic local interaction scheme  (\ref{eq:deform:4mom:generic:int}), the local periodicity of the field $T^{\mu}(x)$  transforms as a ``tangent'', \cite{Dolce:tune,Kenyon:1990fx} 4-vector (\ref{eq:deform:4period:generic:int}) (\emph{i.e}  $T^{\mu}\propto \partial x^{\mu} = e^{\mu}_{a}(x) \partial x^{a}$). In fact,   $\bar p_{\mu}  \propto \partial /\partial x^{\mu}$ and the periodicity are fixed by  (\ref{eq:4D:debroglie:period}). 
The boundary of the theory transforms as an ordinary 4-vector $x'^{\mu}(X) \simeq \Lambda_{a}^{\mu}(x)_{x=X} x^{a}$ where $ e^{a}_{\mu}(x) \simeq \partial x^{a} / \partial x'^{\mu}$, see \cite{Dolce:tune}. In this paper we investigate the case of ``smooth'' interactions, \ie the approximation  $ \Lambda^{a}_{\mu}(x) \sim e^{a}_{\mu}(x)$. \label{footnote:tangent:4D}
} 4-vector \cite{Kenyon:1990fx} (``instantaneous'' de Broglie 4-periodicity).
 Thus it is  possible  to introduce the useful notation $\bar p_{\mu} c = h/ T^{\mu}$, so that the relativistic dispersion relation 
\begin{equation}
\bar{M}^{2}c^{2}=\bar{p}^{\mu}\bar{p}_{\mu}\,,\label{relat:constr:4p}
\end{equation}
can be geometrically interpreted  as the following relativistic constraint on the (``instantaneous'') de Broglie
4-periodicity,  
\begin{equation}
\frac{1}{T_{\tau}^{2}}=\frac{1}{T_{\mu}}\frac{1}{T^{\mu}}\,.\label{relat:constr:4T}
\end{equation}
In the massless case, $\bar M \equiv 0$, \ie infinite proper-time periodicity  $T_\tau \equiv \infty$ or ``frozen'' de Broglie ``internal clock'', we say that  the spatial and temporal periodicities are conformally modulated, $c^{2} T_{t}^{2}(\mathbf{\bar{p}}) = \lambda_{x}^{2}(\mathbf{\bar{p}})$ .

Another important property of the formalism of field theory in compact 4D is that the de Broglie periodicity is encoded on the boundary of the theory  through PBCs. In the cases investigated in this paper the      compactification 4-length can be approximated with the de Broglie 4-periodicity, but this is not true in general, see  \cite{Dolce:tune}  and footnote.(\ref{footnote:tangent:4D}).  Thus the  compactification 4-length   $T^{\mu}$ of (\ref{generic:actin:comp4D}) is dynamically fixed by the 4-momentum (\ref{eq:4D:debroglie:period}). 
As a consequence, in sec.(\ref{Geometrodynamics:boundary}) we will describe relativistic  interactions  in terms of local and retarded deformations of the compact 4D of the theory, \ie in terms of boundary geometrodynamics with interesting analogies with the holographic principle \cite{'tHooft:1993gx,Susskind:1994vu}.

 As we will discuss in more detail, the compact world-line parameter of massive fields in our theory plays a role similar to the compact world-sheet parameter of string theory, so that (\ref{generic:actin:comp4D}) can be regarded as the action of a simple bosonic string  and the de Broglie ``periodic phenomenon'' as a string vibrating in compact 4D. 
 
 \subsection{Mode expansion}

 In the first part of the paper we will only consider the free case, \ie  persistent 4-periodicity $T^\mu$. An isolated ``periodic phenomenon'' in a given reference frame is in fact represented by the action (\ref{generic:actin:comp4D}) with persistent compactification length $T^{\mu}$. 
  The bosonic field  solution (\ref{eq:cond:4D:period})  with persistent periodicity $T^{\mu}$ can be expanded in the harmonics modes of a ``string'' vibrating in compact 4D or in analogy with  a ``particle in a box''. 
In fact, in a given reference frame denoted by $\bar{\mathbf{p}}$, if we consider only time periodicity  $T_{t}(\bar{\mathbf{p}})=2\pi/\bar{\omega}(\mathbf{\bar{p}})$,
 by discrete Fourier transform we find that the field solution has quantized harmonic angular-frequency spectrum ${\omega_{n}}(\mathbf{\bar{p}})=n\bar{\omega}(\mathbf{\bar{p}}) = n  / T_{t}(\bar{\mathbf{p}})$.  Multiplying by the  Planck constant, this corresponds to the quantized energy spectrum 
  \begin{equation}
E_{n}(\bar{\mathbf{p}})=n\bar{E}(\bar{\mathbf{p}})=n\frac{h}{T_{t}(\bar{\mathbf{p}})}\,.\label{eq:energyspectr:period}
\end{equation} 
Thus the periodic field  (\ref{eq:cond:4D:period}), solution of the action  in compact 4D (\ref{generic:actin:comp4D}), can be written as a quantized tower (wave packet) of   3D energy eigenmodes $\Phi_{n}(\mathbf{x})$, 
 \begin{equation}
\Phi(x)\equiv \sum_{n}\Phi_{n}(x)=\sum_{n}e^{-\frac{i}{\hbar}n\bar{E}(\bar{\mathbf{p}})t}\Phi_{n}(\mathbf{x})\,.\label{eq:4d:periodic:field}
\end{equation}

We will address the quantities related to the fundamental mode  $n=1$ with the bar sign, \emph{e.g.} $\bar \Phi_{1}(x) = \bar \Phi(x)$, $E_{1}(\bar{\mathbf{p}}) = \bar E(\bar{\mathbf{p}})$, and so on.
As is well known from the relativistic Doppler effect  the time periodicity $T_{t}(\bar{\mathbf{p}})$ varies with the reference frame $\bar{\mathbf{p}}$.  The variation of time periodicity is described by (\ref{Lorentz:transf:T}) and must fulfill (\ref{relat:constr:4T}). Thus, from (\ref{relat:constr:4p}) we see that the dispersion relation  of fundamental mode $\bar \Phi(x)$  is actually the dispersion relation  
  $\bar{E}^{2}(\bar{\mathbf{p}})=\bar{\mathbf{p}}^{2}c^{2}+\bar{M}^{2}c^{4} $ of a classical particle with mass $\bar M$. 
By plugging this in (\ref{eq:energyspectr:period}), it is easy to see that the dispersion relation of the quantized harmonic energy spectrum of a free periodic field $\Phi(x)$ 
is \begin{equation}
E_{n}(\bar{\mathbf{p}})=n\bar{E}(\bar{\mathbf{p}})=n\sqrt{{\bar{\mathbf{p}}^{2}c^{2}+\bar{M}^{2}c^{4}}}\,.\label{eq:normal:ordered:spectr}
 \end{equation}
 
It is interesting to note that this is nothing other than the energy spectrum of an ordinary second quantized bosonic fields (after normal ordering). This is the first element of a series  of remarkable correspondences between field theory in compact 4D and ordinary QFT, fully discussed in \cite{Dolce:2009ce,Dolce:tune}. They will be   summarized in sec.(\ref{sec:Correspondence-with-QFT}) for the free case and in sec.(\ref{sub:VXD/QFT--correspondence}) for the interacting case. By using common terminology of  QFT we simply say that $\bar{E}(\bar{\mathbf{p}})$ and $\bar{\mathbf{p}}$ are  the energy and the momentum of the 4D field $\Phi(x)$, respectively.

Besides the intrinsic time periodicity we must also consider the induced de Broglie spatial periodicity of our field solution $\Phi(x)$ of topology $\mathbb S^1$. Denoting by $\lambda_{x}$  modulo of the spatial wavelength, the resulting harmonic quantization of the momentum spectrum  is
\begin{equation}
|\mathbf{p}_{n}|=n|\mathbf{\bar{p}}|=n \frac{{h}}{\lambda_{x}(\bar{\mathbf{p}})} \,.\label{eq:quant.wavenum:def}
\end{equation}

Through the Planck constant we can say that the energy and the momentum are the ``physical conjugate'' variables of the  temporal and of the spatial coordinate, respectively.  Therefore, through discrete Fourier transformation, the  intrinsic de Broglie 4-periodicity of $\Phi(x)$ implies a discretization (quantization) of the energy-momentum spectrum: $p_{n \mu}= n \bar p_{\mu}$. Such a ``periodic phenomenon'' has fundamental topology $\mathbb S^1$, so that its harmonic spectrum turns out to be written in terms of the single (quantum) number $n$. 

The simple topology $\mathbb S^{1}$ investigated so far, however, must be extended with two other spherical coordinates in isotropic or topological equivalent systems. In case of spherical symmetry in fact the ``periodic phenomenon'' describing the particle can be parameterized in terms of the angles $\theta \rightarrow \theta +  \pi$ and $\varphi \rightarrow \varphi + 2  \pi$, \cite{Dolce:2009cev4}. As a consequence of the resulting topology $\mathbb S^{1}\otimes \mathbb S^{2}$, in addition to the quantization of the 4-momentum labeled by $n$, we obtain the ordinary quantization of the angular momentum described in terms of two additional quantum numbers,  typically denoted by $\{l,m\}$.  For the sake of simplicity in this paper we will not consider this further expansion of the field $\Phi(x)$ in spherical harmonics or their deformations. 

In the rest frame, the quantized energy spectrum (\ref{eq:normal:ordered:spectr}) of a periodic field $\Phi(x)$, or equivalently to the energy spectrum of a second quantized KG field, yields a quantized mass spectrum $M_{n}   \equiv {E_{n}(0)}/{c^{2}}$ such that
\begin{equation}
M_{n}  = n\bar{M} =n\frac{h}{c \lambda_{s}} \,.\label{eq:KK:energy:spectr}
\end{equation}
According to (\ref{Compton:mass:period}), $\lambda_{s}$ is nothing other but the Compton wavelength of the field of mass $\bar M$.
The mass is the ``physical conjugate'' of the world-line parameter. As for the semi-classical quantization of the energy-momentum described above, in this case the discretized mass spectrum is a \emph{direct} consequence of the intrinsic proper-time periodicity of the field  $\Phi(x)$. In sec.(\ref{sec:Virtual-Extra-Dimension}) we will see that this result, formally equivalent to the KK mass tower, is part of a more general dualism between periodic fields and XD fields.

   In ordinary QFT  we can imagine to fix the value of the field at
the boundaries (\eg through SBCs) in order to select a particular single mode $\Phi_{KG}(x)$,  solution of the Klein-Gordon (KG) equation, and then impose commutation relations. The difference with the
ordinary approach to field theory (where BCs have however a marginal role in practical computations) is that in (\ref{generic:actin:comp4D:transf}) we fix a particular solution of the corresponding relativistic differential system
by imposing PBCs rather than SBCs.  As shown in \cite{Dolce:tune} the formalism of field in compact 4D is useful to study classical (\textit{i.e.} non-quantum) relativistic fields.  
In fact, the fundamental mode $\bar \Phi(x) \propto e^{-\frac{i}{\hbar} \bar p_\mu x^{\mu}}$  of the periodic field solution $\Phi(x)$ can be locally matched to a corresponding  single KG mode $\Phi_{KG}(x) = \bar \Phi(x)$  of an ordinary non-quantized  KG field  of mass $M_{KG}=\bar M$ and energy ${E}_{KG}(\bar{\mathbf{p}}) =\bar{E}(\bar{\mathbf{p}})$ (\textit{i.e.} to a single de Broglie matter wave).  Thus, the fundamental mode $\bar \Phi(x)$, as  $\Phi_{KG}(x)$,  describes the relativistic behavior of a classical particle.  This means that ordinary QFT  can be retrieved by the second quantization of the fundamental mode $\bar \Phi$, neglecting all the higher modes (for instance by imposing SBCs instead of PBCs), \cite{Dolce:tune} 

Nevertheless, as we will see in the next section, the assumption of PBCs can be used as a semi-classical quantization condition. In fact, if all the harmonics are considered,  a de Broglie ``periodic phenomenon'' is described as an harmonic wave-packet whose evolution has remarkable formal overlaps with an ordinary second quantized  field, \cite{Dolce:2009ce}.  In case of interaction however the periodicity of the corresponding harmonic wave-packet must be locally modulated to encode the corresponding variations of kinematical state. In sec.(\ref{sec:Towards-Interactions}), the strategy to  describe local  interactions will be that of modulating locally the spatial and temporal intrinsic periodicities of the field $\Phi(x)$ (instead of using creation and annihilation operators). This will be realized by associating to every local space-time point $X$ an action with locally deformed compactification length $T^{\mu}(X)$.

\subsection*{Comments  and  Outlooks}

In defense of the scientific method, de Broglie said ``\emph{this hypothesis [of periodic phenomenon] is at the base of our theory: it is worth as much, like all hypothesis, as can be deduced from its consequences}'', \cite{Broglie:1924}. In this spirit, similar to recent papers \cite{Dolce:2009ce,Dolce:tune}, our study  wants to explore  the validity and the consistency, in applications of modern physics, of such a description of elementary particles as intrinsically periodic phenomena. 

In our formalism the intrinsic periodicity is realized in terms of field theory in compact 4D through PBCs. We have already seen elements of the formal consistency of this theory with relativity, such as its covariance.  Further fundamental elements of its consistency with QM and GR will be discussed in sec.(\ref{sec:Correspondence-with-QFT}) and sec.(\ref{sec:Towards-Interactions}), respectively.  Essentially we find out that, as long as we consider the relativistic modulations of periodicity, our formulation  turns out to be a consistent quantum-relativistic theory. Moreover, it is important to bear in mind that, from a formal point of view, SR---and  GR as we will see later---defines the differential properties of space-time without giving any particular restriction about the BCs whose only requirement is to fulfill the variational principle.  Indeed,  \cite{Dolce:2009ce,Dolce:tune},  such a dynamical  description of ``periodic phenomena'' is achieved by applying consistently the variational principle at the geometrodynamical boundary of a relativistic wave theory.   On the other hand, it is well known that BCs play a fundamental role in QM as we will discuss in the next section.

Besides the purely mathematical aspects of the theory, in this section, as well as in similar sections throughout the whole paper, we want to provide conceptual arguments to interpret the formal consistency of our results. In particular here we want to justify the formalism of compact 4D  in the context of SR and undulatory mechanics.  

Neglecting for a moment the mode expansion of a ``periodic phenomenon'', to figure out the possibility of a description of elementary particles in terms of angular variables (such as the real coordinates appearing in the phase of waves) we may, for instance, follow few simple logical steps: 
\begin{enumerate}
\item[\emph{i})]  According to classical-relativistic mechanics, every elementary particle, in a given local reference frame, is  characterized by an energy $\bar E(\bar{\mathbf{p}})$ and momentum $\bar{\mathbf{p}}$. In the free case the particle is in an inertial frame $\bar{\mathbf{p}}$ and its 4-momentum $\bar p_\mu $ is constant (\ie Newton's law of inertia).
\item[\emph{ii})] In undulatory mechanics every elementary isolated particle has associated intrinsic temporal and spatial de Broglie periodicity $T_t (\bar{\mathbf{p}})=h/\bar E (\bar{\mathbf{p}}) $ and $\lambda^i=h/p_i$ (de Broglie hypothesis or wave-particle duality). The resulting local 4-periodicity $T^\mu$ depends on the local reference frame and it is  persistent for isolated particles. 
\item[\emph{iii})] In the atomistic description characterizing modern physics, elementary particles are the fundamental constituents of every system in nature. These elementary particles can interact exchanging 4-momentum $\bar p_\mu $ and thus modulating their 4-periodicity $T^\mu$, according to the relativistic laws. 
\end{enumerate}
\emph{Hence, the logical combination of these scientific truths naturally implies the possibility of a formulation of physical systems in terms of intrinsically ``periodic phenomena'' representing the elementary particles}. This  justifies  the good properties of our formalism. In few words, the consistency of this description is already implicit in the ordinary undulary description of elementary particle (wave-particle duality) at the base of modern QFT.

 In particular, we may imagine a system composed by a single free elementary particle.  Since in the undulatory description such an isolated particle has persistent periodicity, it can be imagined as a pendulum in the vacuum. Indeed, the periodic oscillations of this system can be parameterized by an angular variable, \emph{e.g.} by a phasor or a wave. For instance, as well known, a free classical-relativistic particle can be described by a corresponding KG mode $\Phi_{KG}$. Similarly, adding  more isolated particles, the system can be described by a set of angular variables parameterizing  every elementary ``periodic phenomenon''. 

For the sake of simplicity in this conceptual digression we consider only the time variable.  In the international system (SI) the unit of time ``second'' is defined as the duration of 9,192,631,770 characteristic cycles of the Cs atom.  Indeed, time  can be only defined by counting the cycles of a phenomenon which is \emph{supposed} to be periodic in order to guarantee the invariance of the unit of time. We may for instance think to Galileo's experiment of the pendulum isochronism in the Pisa dome\footnote{For some aspects this can be regarded as the beginning of modern physics since it allowed a sufficient accuracy in the measurement of time to test theory of dynamics.} or to Einstein's definition of a relativistic clock in which ``all that happens in a  given period is identical with all that happens in an arbitrary period'', see sec.(\ref{Sec:Compact:spacetime}). Thus, an isolated elementary particle, having persistent periodicity as a pendulum in the vacuum, can be regarded as a reference clock, the so-called de Broglie ``internal clock'', \cite{1996FoPhL,2008FoPh...38..659C}.  

For an isolated system (universe) composed by a single ``periodic phenomenon'', as for an isolated pendulum in the vacuum, time can be regarded as a cyclic variable since ``all that happens in a  given period is identical with all that happens in an arbitrary period''. Being that  the system is isolated, there are no external variations of energy  distinguishing a given period from any other. Since the whole physical information is contained in a single period we say, by using a terminology typical of XD theories, that the time of such an isolated system is compactified on a circle $\mathbb S^1$ (cyclic universe).     If we parameterize its oscillation with  an external time axis $t \in \mathbb R$, for instance defined by the ``ticks'' of the Cs clock,  we see that the time parameter enters into the equations of motion as an angular variable with periodicity fixed by the energy of the system as in a wave function. The external time axis can be also regarded as the time of a reference system of infinite periodicity or whose periodicity is many orders of magnitude bigger than the time scales under investigation. In this way  we find out that the angular variables of different ``periodic phenomena'' can be parameterized by the same parameter $t$. Thus, in a generic interval $t \in (t_i, t_f)$ on such an external axis,   a ``periodic phenomenon'' $\Phi(\mathbf{x},t)$ assumes particular initial and final values  $\Phi(\mathbf{x},t_i)$ and $\Phi(\mathbf{x},t_f)$ during its evolution. This means that a ``periodic phenomenon'' is not localized in a particular temporal region.  However, the evolution is defined modulo periodic translations, as we will see this  yields an intuitive interpretation of quantum evolution of the system in terms of the FPI.   On the other hand, see \cite{Dolce:2009ce}, in this periodic system, if we imagine turning on a source of energy, say at $t_i$,  the energy propagates according to the retarded relativistic potential (the theory is in fact based upon relativistic differential equations). After a given delay, say at time $t_f$, this induces a variation of periodic regime to the particle, depending on the amount of energy exchanged. As a result, the system passes from a periodic regime to another periodic regime so that it is possible to establish a before and an after with respect to this retarded event in time (in this case all that happens in a  given period is \emph{not} identical with all that happens in an arbitrary period). Hence we can intuitively understand relativistic causality and time ordering of events in terms of modulations of periodicity of the elementary particles' internal clocks  \cite{Dolce:2009ce}. 

In a system composed by more non-interacting particles, the same external time parameter appears in the EoMs of every ``periodic phenomenon'' as an angular variable with different periodicity, depending on its kinematical state. 
However, such a non-elementary system in general has \emph{no} periodic dynamics. As the ratio of periodicities does not necessarily form a rational number, the resulting evolution is  indeed an ergodic evolution \footnote{A system can be regarded as elementary as long as our resolution does not allows us to resolve its possible egodics dynamics.}.

In particular, every isolated elementary particle can be regarded as a reference clock so that, in principle, its ``ticks'' can be  used to define the external time axis similarly to the Cs clock\footnote{For instance, this would yield a remarkable improvement in the experimental resolution of time, if only it were possible to count the ``ticks'', say, of an electron without perturbing its periodicity.}. 
If we also consider relativistic effects, every external observer or elementary particle in the system describes a different combination of the phases of the de Broglie internal clocks and thus a different ``present'', depending on their relativistic kinematic state.  This is nothing other than the ordinary relativistic description of simultaneity.

 Every value (instant in time) of an external relativistic time axis $t \in \mathbb R$ can be characterized by a unique combination of the ``ticks'' of the particle ``internal clocks'', as in a stopwatch or in a calendar\footnote{In everyday life we fix events in time in terms of reference periods of years, months, days, hours, minutes, seconds. These reference cycles are conventionally rational each other,  in particular in a sexagesimal base. But  they need regular  adjustments since they mimic natural cycles (\eg Moon and Earth rotations) which are not rational to each other.}. Thus, as noted in previous publications, \cite{Dolce:2009ce,Dolce:tune,Dolce:2009cev4,Dolce:QTRF5,Dolce:2010ij,Dolce:Dice,Dolce:FQXi, Dolce:Cyclic}, this gives rise to a possible scenario in which the external time axis can be in principle eliminated in describing events in time.  
To complete the physical consistency of this picture we must however consider the role of the mediator of interactions among particles, and the consequent relational description of events. As mentioned at the beginning of sec.(\ref{Covariant:Notation}), massive particles, expect neutrinos, have typically extremely fast periodicity. For instance the electron intrinsic time periodicity  is faster than or equal to its proper-time compactification length $T_{\tau } = 8.093299724 \pm 11 \times 10^{-21} s$. On the other hand, the time periodicity of a massless particle such a photon or graviton can vary from zero to infinity. Since the world-line compactification length, say, of photons is infinite  we say that light has a ``frozen proper-time internal clock'' $T_{\tau } \equiv \infty $. 
In a relational description of the elementary particles, the long temporal and spatial compactification lengths of light provide the long space-time scale structure of the system. They can be regarded as the reference temporal (and spatial) axis of ordinary relativity, allowing a common parameterization of the elementary cyclic phenomena;  such long space-time scales with respect to the typical periodicities of the matter fields can be used as a reference upon which the ordinary relativistic structure of space-time can be built. In particular the ``frozen'' clock of light or of gravity plays the role of the emphatically non-cyclic world-line in relativity.  We will come back to this point in the next sections. 

An interacting particle will be formally described by modulating the de Broglie 4-periodicity. This description can also be extended to a single KG mode $\Phi_{KG}$, as pointed out in \cite{Dolce:tune}. 
 In addition to this, if we imagine to switch on interactions among particles, the modulations of periodicity associated with the exchange of energy-momentum yield  very chaotic evolutions of the system  --- thus the theory does not necessarily imply ordinary cyclic cosmology. 

Summarizing, this description in which every  elementary particle can be regarded as a Einstein relativistic reference clock,  not only is fully consistent with the special relativity, but the local nature of relativistic time turns out to be enforced. This provides interesting new elements to address the notion of time in physics.  We will not discuss further the problem of the flow of time and time symmetry since these go beyond the scope of the paper (this would involve interesting philosophical and historical digressions).  Some more detail  about this point has been recently given in  \cite{Dolce:2009ce,Dolce:2009cev4,Dolce:Dice,Dolce:FQXi,Dolce:Cyclic}. Also we mention that the possibility of an effective description of the arrow of time and of the quantum behavior of elementary particles in terms of ergodics dynamics (for instance associated to curled-up extra dimensions) has been proposed in \cite{Elze:2003tb}. 

In this paper we will rather discuss another important physical and conceptual motivation of our assumption of intrinsic periodicity, that is to say the formal correspondence with QM, in both the canonical and Feynman formulation, arising from the assumption of PBCs at the ends of compact 4D. In fact, our assumption naturally incorporates the undulatory nature of quantum particles, generalizing the central role that BCs had in the  ``old'' formulation of QM.

\section{Correspondence to free QFT \label{sec:Correspondence-with-QFT}}

We now want to study in more in detail the mechanics of the periodic solution of our field theory in compact 4D. We assume persistent 4-periodicity $T^{\mu}$, so that  the results of this section concern essentially free particles. 
We will find that  the mechanics of our periodic fields have remarkable formal correspondences with  ordinary QM \cite{Dolce:2009ce}.
In this paper, as exemplification, we mainly consider the correspondence with the FPI formulation of QM; more evidence is given in \cite{Dolce:2009ce,Dolce:tune}. For the sake of simplicity here we  assume
a single spatial dimension denoted by $\mathrm{x}$,  avoiding expansion in spherical harmonics. 

The periodic scalar
field (\ref{eq:4d:periodic:field}) with dispersion relation 
$E_{n}(\mathbf{\bar{p}})$ in (\ref{eq:normal:ordered:spectr})
 can  be written with the following notation \cite{Kapusta:1989tk}\begin{eqnarray}
\Phi(\mathrm{x},t)  = \sum_{n}A_{n}a_{n}\phi_{n}(\mathrm{x},t) 
= \sum_{n}A_{n}a_{n}e^{-\frac{i}{\hbar}(E_{n}t-p_{n}\mathrm{x})}~. \label{hilbert:H} \end{eqnarray}
The normalization factors $A_{n}$ are fixed for instance by choosing
the inner product induced by the conservation of the ``charge density'' \footnote{The purely translational zero mode $n=0$ can be regarded as unphysical. }
\begin{eqnarray}
\left\langle f | g \right\rangle _{Q} & = & \int_{0}^{\lambda_{\mathrm{x}}}d \mathrm{x} f^{*}(\mathrm{x},t)\frac{i\stackrel{\leftrightarrow}{\partial_{t}}}{\hbar c^{2}}g(\mathrm{x},t)~,\label{Q:prod}\end{eqnarray}
 where $A\stackrel{\leftrightarrow}{\partial_{t}}B=A(\partial_{t}B)-(\partial_{t}A)B$, so that $A_{n}  =  \sqrt{\frac{\hbar c^{2}}{2\omega_{n}\lambda_{\mathrm{x}}}}$.
 Similarly to ordinary field theory, the coefficients $a_{n}$ of the Fourier expansion  are given
by $
a_{n}=\left\langle f_{n}(\mathrm{x},t)|\Phi_{n}(\mathrm{x},t)\right\rangle _{Q}
$, where $f_{n}(\mathrm{x},t) = A_{n} \phi(\mathrm{x},t)$.

A periodic field is essentially 
a sum over the harmonic modes of a vibrating classical
string. This actually is the typical classical system which can be described in a Hilbert space.  In fact, the energy eigenmodes $\phi_{n}(\mathrm{x})$ form
a complete set with respect to the inner product (``non-relativistic limit'' of  $\left\langle \; | \; \right\rangle _{Q} $) 
\begin{equation}
\left\langle \phi|\chi\right\rangle \equiv\int_{0}^{\lambda_{\mathrm{x}}}\frac{{d\mathrm{x}}}{{\lambda_{\mathrm{x}}}}\phi^{*}(\mathrm{x})\chi(\mathrm{x})\,.\label{Hilbert:innerprod}\end{equation}
The energy eigenmodes  can be used to define Hilbert eigenstates
\begin{equation}
\left\langle {\mathrm{x}}|\phi_{n}\right\rangle \equiv\frac{
{\phi_{n}({\mathrm{x}})}
}{{\sqrt{\lambda_{\mathrm{x}}}}}
\,.\label{Hilbert:eigenstates}\end{equation}
Note that the integral over a single spatial period $\lambda_{\mathrm{x}}$ can be extended  to an arbitrary large (or infinite) number $N_{T} \in \mathbb N$ of periods:     $\int_{0}^{\lambda_{\mathrm{x}}}\frac{{d\mathrm{x}}}{{\lambda_{\mathrm{x}}}}\rightarrow\int_{V_{\mathrm{x}}}\frac{d\mathrm{x}}{V_{\mathrm{x}}}$ with $V_{\mathrm{x}}= N_{T} \lambda_{\mathrm{x}}$. 

The evolution along the compact time, described by the so-called
bulk EoMs $(\partial_{t}^{2}+\omega_{n}^{2})\phi_{n}({\mathrm{x}},t)=0$, can be reduced to first order differential equations \cite{Carena:2002me}, so that
 \begin{equation}
i\hbar\partial_{t}\phi_{n}({\mathrm{x}},t)=E_{n}\phi_{n}({\mathrm{x}},t).\label{eq:pre:schrod:eqs}\end{equation}
 This set of differential equations 
 corresponds
to the ordinary Schr\"odinger equation of QM. In fact, \cite{Nielsen:2006vc},
from the Hilbert eigenstates (\ref{Hilbert:eigenstates}), 
 it is possible to formally define a Hamiltonian operator $\mathcal{H}$ such that
\begin{equation}
\mathcal{H}\left|\phi_{n}\right\rangle _{}\equiv E_{n}\left|\phi_{n}\right\rangle \,.\label{eq:hamilt:def:gen:state}\end{equation}
Similarly, the momentum operator $\mathcal{P}$ can be defined as
\begin{equation}
\mathcal{P}\left|\phi_{n}\right\rangle \equiv-p_{n}\left|\phi_{n}\right\rangle \,.\label{eq:moment:def:gen:state}\end{equation}
A  Hilbert state $|\phi\rangle=|\phi(t')\rangle$ describing a generic periodic field at time $t'$,
is a generic superposition of Hilbert eigenstates\begin{equation}
|\phi\rangle\equiv\sum_{n}a_{n}|\phi_{n}\rangle\,.\end{equation}
Therefore, from the EoMs (\ref{eq:pre:schrod:eqs}), we find 
that the time evolution  is actually described by the familiar Schr\"odinger equation of ordinary QM, \begin{equation}
i\hbar\partial_{t}|\phi(t)\rangle=\mathcal{H}|\phi(t)\rangle.\end{equation}

 From (\ref{hilbert:H}) we see that the time evolution of the generic Hilbert state $|\phi\rangle$
is also described by   the  exponential operator 
\begin{equation}
\mathcal{U}(t';t)=e^{{-\frac{{i}}{\hbar}\mathcal{H}(t-t')}}\,.\end{equation}
Moreover this time evolution is Markovian (unitary), and can be
written as product of $N$ infinitesimal evolutions,
 \begin{equation}
\mathcal{U}(t'';t')=\prod_{m=0}^{N-1}\mathcal{U}(t'+t_{m+1};t'+t_{m}-\Delta t)\,,\label{eq:marct:time:operator}
\end{equation}
where $N \Delta t =t''-t'~$.

Without any further assumption than intrinsic periodicity
 we have obtained all the elements necessary to build the ordinary Feynman Path
Integral (FPI) formulation of QM. In fact, if we  plug the completeness relation of the energy eigenmodes in between the
elementary Markovian evolutions, we obtain the finite evolution
\begin{eqnarray}
\mathcal{Z} 
&=&\int_{V_{\mathrm{x}}}\left(\prod_{m=1}^{N-1}\frac{d\mathrm{x}_{m}}{V_{\mathrm{x}}}\right)\mathcal U({\mathrm{x}}'',t";{\mathrm{x}}_{N-1},t_{N-1})\times\dots\nn\label{int:evol:oper}\\
 &  & \dots\times \mathcal U({\mathrm{x}}_{2},t_{2};{\mathrm{x}}_{1},t_{1}) \mathcal U({\mathrm{x}}_{1},t_{1};{\mathrm{x}}',t')\,.\label{mark:element:PI}
 \end{eqnarray}
From the notation introduced  so far  the elementary 4D
evolutions can be written in the following way 
\begin{eqnarray}
{\mathcal U(\mathrm{x}_{m+1},t_{m+1};\mathrm{x}_{m},t_{m})} & = & \sum_{n_{m}}{e}^{-\frac{i}{\hbar}(E_{n_{m}}\Delta t_{m}-{p}_{n_{m}}\Delta{\mathrm{x}}_{m})}\nn\\
 & = & \left\langle \phi\right|e^{-\frac{i}{\hbar}(\mathcal{H}\Delta t_{m}-\mathcal{P}\Delta \mathrm{x}_{m})}\left|\phi\right\rangle,\label{elemet:phase:space}
 \end{eqnarray}
 where $\Delta \mathrm{x}_{m}=\mathrm{x}_{m+1}-\mathrm{x}_{m}$ and $\Delta t_{m}=t_{m+1}-t_{m}$. As in the usual FPI formulation we are assuming on-shell elementary
4D evolutions \cite{Feynman:1942us}.
Therefore, \cite{Dolce:2009cev4}, we have  obtained that the classical evolution of our system with intrinsic periodicity is formally described by the ordinary FPI, which in the phase-space formulation is   
\begin{equation}
\mathcal{Z}=\!\!\lim_{N\rightarrow\infty}\int_{V_{\mathrm{x}}}\!\!\left(\prod_{m=1}^{N-1}\frac{d\mathrm{x}_{m}}{V_{\mathrm{x}}}\!\right)\!\!\prod_{m=0}^{N-1}\!\left\langle \phi\right|e^{-\frac{i}{\hbar}(\mathcal{H}\Delta t_{m}-\mathcal{P}\Delta \mathrm{x}_{m})}\left|\phi\right\rangle. \label{periodic:path.integr:Oper:Fey}
\end{equation}
This remarkable result has been obtained in a semi-classical way, without any further assumption than intrinsic 4-periodicity.

In complete analogy with  the ordinary Feynman formulation,
(\ref{periodic:path.integr:Oper:Fey}) can be expressed in configuration space by means of the action 
 \begin{equation}
{\mathcal{S}}_{cl}(t_{b},t_{a})\equiv\int_{t_{a}}^{t_{b}}dt {L}_{cl}\label{class:pt:action}\,.
\end{equation}
This is nothing other than  the action of the  free classical particle associated with our ``periodic phenomenon''. In fact it is defined by the Lagrangian  
\begin{equation}
 {L}_{cl}\equiv \mathcal{P} \dot \mathrm{x}_{m } - \mathcal{H} \label{class:pt:lagrang}
\,,\end{equation}
Thus\footnote{As in the ordinary formulation of the FPI, to pass from the phase-space formulation to the configuration space description
  we have performed the sum over the momentum spectrum of the Hilbert space \cite {Dolce:2009cev4}.},
the infinitesimal evolutions can be written as   \begin{eqnarray}
\left\langle \phi\right|e^{-\frac{i}{\hbar}(\mathcal{H}\Delta t_{m}-\mathcal{P}\Delta \mathrm{x}_{m})}\left|\phi\right\rangle 
 & = & e^{\frac{i}{\hbar}{\mathcal{S}}_{cl}(t_{m+1},t_{m})}\,.\label{eq:elem:xt:evol:2}\end{eqnarray}

Finally, by using the notation of the functional measure\footnote{$\int_{V_{\mathrm{x}}} {\mathcal{D}\mathrm{x}} = \lim_{N\rightarrow\infty} \int_{V_{\mathrm{x}}} \prod_{m=1}^{N-1}\frac{d\mathrm{x}_{m}}{V_{\mathrm{x}}}$.},  the FPI  (\ref{periodic:path.integr:Oper:Fey})
 in  configuration space describing the classical evolution of a classical ``periodic phenomenon'' is \begin{equation}
\mathcal{Z}=\int_{V_{\mathrm{x}}} {\mathcal{D}\mathrm{x}}  e^{\frac{i}{\hbar}  \mathcal{S}_{cl}(t_{f},t_{i})}\,.\label{eq:Feynman:Path:Integral}\end{equation}

It is important to note that  this demonstration, being based on the composition of elementary evolutions,  can be formally generalized to a non-homogeneous Hamiltonian $\mathcal{H}'$ through the formal substitution $\mathcal{H} \rightarrow \mathcal{H}'$. 
As shown in \cite{Dolce:tune} and as we will summarize in sec.(\ref{sub:VXD/QFT--correspondence}), in case of interaction the free action $\mathcal{S}_{cl}$ of the FPI (\ref{eq:Feynman:Path:Integral}) will be substituted with the action of the corresponding interacting classical particle $\mathcal{S}_{cl}\rightarrow \mathcal{S}'_{cl}$.

We also note that, \cite{Dolce:2009ce}, by evaluating the expectation value associated to the Hilbert space (\ref{Hilbert:innerprod}-\ref{Hilbert:eigenstates}) of the observable $\partial_{\mathrm{x}} F(\mathrm{x})$, and by integrating by parts,   we find
	   \begin{equation}
\left\langle \phi_{f} | \partial_{\mathrm{x}} \mathcal{F}(\mathrm{x})  |\phi_{i}\right\rangle = \frac{i}{\hbar} \left\langle \phi_{f} | \mathcal{P} \mathcal{F}(\mathrm{x}) - \mathcal{F}(\mathrm{x}) \mathcal{P}  |\phi_{i} \right\rangle\,.
\end{equation}
In fact the boundary terms vanish because of the periodicity of the spatial coordinate.
	  By assuming that the observable is such that $\mathcal{F}(\mathrm{x})\equiv \mathrm{x}$ \cite{Feynman:1942us}, the above equation  for generic   initial and final Hilbert  states $ |\phi_{i}\rangle $ and $ |\phi_{f}\rangle$,  is nothing other than the commutation relation  $
[\mathrm{x},\mathcal{P}]=i  \hbar 
$ 
of ordinary QM --- more in general $ [\mathcal{F}(\mathrm{x}), \mathcal{P} ] = i \hbar \partial_{\mathrm{x}} \mathcal{F}(\mathrm{x})$. The commutation relations can be regarded as implicit in this theory. With this result we have checked the correspondence with canonical QM. Note that a similar demonstration was used by Feynman in \cite{Feynman:1942us} to show that the FPI formulation of QM is equivalent to the canonical formulation.  

\subsection*{Comments and Outlooks}

The correspondence to ordinary relativistic mechanics obtained above is one of the main motivations for our formalism based on compact 4D and PBCs. 
The FPI obtained in (\ref{eq:Feynman:Path:Integral}) can be intuitively,
as well as formally and graphically \cite{Dolce:2009ce,Dolce:tune,Dolce:2009cev4,Dolce:Cyclic}, interpreted in a semi-classical way as the consequence of the topology $\mathbb S^1$ associated to the de Broglie ``periodic phenomenon''. In fact in such a cylindrical geometry,
because of the invariance under 4-periodic translations (\ref{eq:cond:4D:period}),
the initial and final configurations of the field $\Phi({x}_{i})$ and
$\Phi(x_{f})$ can be reached by an infinite set of periodic classical paths characterized
by different winding numbers. In other words, because of the PBCs a field in compact
4D can self-interfere, and the resulting evolution turns out to be the one prescribed by ordinary QM.

The reader interested in more correspondences between fields in compact 4D
and ordinary QFT may refer to \cite{Dolce:2009ce,Dolce:tune}. For instance, the assumption of intrinsic periodicity in the free case leads to the harmonic spectrum $p_{n \mu} T^{\mu}= n \bar p_{ \mu} T^{\mu} = n h$, see (\ref{eq:normal:ordered:spectr}). More in general, as shown in \cite{Dolce:tune}, in case of interaction the quantized spectrum associated with the corresponding modulation  of periodicity 
turns out to be given by the Bohr-Sommerfeld condition $\oint p_{n \mu}(x) d x^{\mu} = h (n + v) $ (see also \cite{Solovev:2011}, the so-called Morse index $v$ can be retrieved by assuming a twist factor in the PBCs). Similarly to the
Bohr hydrogen atom or to a particle in a box, the allowed orbits and energy levels are those with an integer number of cycles \cite{Dolce:2009ce,Dolce:2009cev4,Solovev:2011}. 
This approach has been tested in the description of several quantum phenomena \cite{Dolce:2009cev4} such as black-body radiation, double slit experiment, wave particle duality, etc, as well as in applications of non-relativistic quantum mechanics and condensed matter  such as superconductivity, Schr\"odinger problems, atomic physics, etc, \cite{Dolce:2009cev4,Dolce:superc,Solovev:2011}.

It is instructive to interpret the black-body radiation in terms of ``periodic phenomena''. In this case it is natural to assume that the population $a_n$ of the $n^{th}$ energy levels is fixed by the Boltzmann  distribution. For the IR components of the radiation, \ie massless periodic phenomena with long periodicity, the PBCs can be neglected and the energy spectrum can be approximated to a continuum (classical limit) since many energy levels are populated (we may say that the thermal noise destroys the intrinsic periodicity in a sort of decoherence). For  the UV components, however,  the fundamental energy $\bar E$ is big with respect to the thermal energy so that only a few energy levels can be populated. That is to say, these modes have very short periodicity (not destroyed by the thermal noise), so that the PBCs are important and there is a manifest quantization of the energy spectrum (quantum limit). The UV catastrophe is avoided according to Planck.    

For massive particles, in the non-relativistic limit $\bar p \ll \bar M c$, only the fundamental energy level of a periodic phenomenon is largely populated as the gap between the energy levels goes to infinity, $\bar M \rightarrow \infty$ \footnote{In the non-relativistic limit or when stopped in a detector, we say that the wavefunction collapses to the ground state . In this limit the inner product (\ref{Q:prod}) describing the relativistic probability to find a charged particle must be replaced by (\ref{Hilbert:innerprod}) (Born rule), \cite{Dolce:2009ce}. }. The cyclic field can therefore be approximated as  $ \Phi(x) \sim \exp[-i \frac{\bar M c^{2}}{\hbar} t + i \frac{\bar M}{\hbar} \frac{\mathbf x^{2}}{2 t}]$, see \cite{Dolce:2009ce}. Neglecting the proper de Broglie ``internal clock'' (first term), it is possible to see by plotting the modulo square, that the wavefunction of a massive ``periodic phenomenon'' is centered along the path of the corresponding classical particle and its width is smaller than the Compton wavelength $\lambda_s$. Thus, in the non-relativistic limit, the Dirac delta distribution is reproduced  as in the usual Feynman description. Similarly, the spatial compactification length tends to infinity whereas the time compatification tends to zero, so that in the non-relativistic limit we have a point-like distribution in $\mathbb R^3$, \ie the ordinary three-dimensional description of a classical particle. Indeed, the corpuscular description therefore arises at high frequencies.        
On the other hand, in the relativistic limit the non-local nature of a massive periodic phenomena can not be neglected (the distribution width is  of the order of the Compton wavelength). This gives an intuitive interpretation of the wave-particle duality and of the double slit experiment. In particular, if probed with high energy or observed with good resolution, more and more energy levels (in general with either positive or negative frequencies) turn out to be excited, \ie more and more harmonics can be resolved. In this way we can figure out that the energy excitations play the role of the quantum excitations of the same fundamental elementary system, so that we have an analogy with the \emph{virtual} particles of ordinary relativistic QM.  This aspect will motivate the terminology ``\emph{virtual} XD'' in describing the cyclic proper-time of an elementary ``periodic phenomenon''.

The assumption of intrinsic periodicity implicitly contains
the Heisenberg uncertain relation of QM. Briefly, the phase of a `` de Broglie clock''  can not be determined and is  defined modulo factor $\pi n$ since only the square of the field has physical meaning, according to (\ref{Hilbert:innerprod})--- this also brings the factor $1/2$ in the resulting uncertain relation. To determine the frequency, and thus the energy
$\bar{E}(\mathbf{\bar{p}})=\hbar \bar{\omega}(\mathbf{\bar{p}})$, with good accuracy $\Delta\bar{E}(\mathbf{\bar{p}})$ we
must count a large number of cycles. That is to say we must observe the system
for a long time $\Delta t(\mathbf{\bar{p}})$ according to the relation $\Delta\bar{E}(\mathbf{\bar{p}})\Delta t(\mathbf{\bar{p}})\gtrsim\hbar/2$.    The simple mathematic demonstration of this relation is given in \cite{Dolce:2009ce}. 

As already pointed out in previous publications, because of
 the extremely fast periodic dynamics typically associated to matter particles, QM can be regarded as an emerging phenomenon. 
The situation is therefore similar to the one considered in recent attempts to interpret QM as an emerging theory, such as in the 't Hooft determinism and in the stroboscopic quantization. According to 't Hooft, there is a ``close relationship between the quantum harmonic oscillator and a particle moving on a circle'' if their periodicity $T_t$  is faster than our resolution in time \cite{'tHooft:2001ar}. This correspondence can be seen in our theory from the fact that the ``periodic phenomenon'' with periodicity $T_t$ describes the quantum behavior of the corresponding KG mode, which corresponds to a quantum harmonic oscillator of the same periodicity. A similar description is given by the ``stroboscopic quantization'' in which  QM and, as already mentioned, the arrow of time are interpreted as  emerging phenomena associated with ergodic dynamics resulting from XDs compactified on a torus \cite{Elze:2003tb}. 

 The  periodicity upper limit  of typical quantum systems, characterized  by electrodynamics,  can be regarded as fixed by the electron (the lighter massive particle except neutrinos) proper-time periodicity $T_{\tau} \sim 10^{-20} s$. Thus, using a Cs clock, $T_{Cs} \sim 10^{-10} s$, to determine the dynamics of such a system is like using a time scale of the order of the age  of the universe ($\sim 10^{10} y$) to investigate annual dynamics. For every ``tick'' of the Cs clock the electron, in its evolution, does an enormous number of cycles, \ie about windings $10^{10}$. The actual experimental time resolution, about $\Delta T_{exp}  \sim 10^{-17}$,  is still too low to resolve such small time scales, though the internal clock of an electron has been observed in a recent interference experiment \cite{2008FoPh...38..659C}.  Indeed the observation of such a fast de Broglie internal clock is similar to the observation of a ``clock under a stroboscopic light''. The results is that at every observation the particle appears to be in an aleatoric phase of its cyclic evolution. As for a dice rolling too fast with respect to our resolution in time, the outcomes can be described only in a statistical way. Loosely speaking, an observer with infinite resolution in time can in principle resolve exactly the underling deterministic dynamics, and would have no fun playing dice (``God doesn't play dice'', Einstein). 
In \cite{Dolce:Dice} we have named a dice rolling with de Broglie frequency as ``de Broglie deterministic dice''.
 Similarly to the deterministic models mentioned above, the formal results of the previous section show that the statistics associated with these extremely fast cyclic behaviors have formal correspondence with ordinary QM. This also suggests that a direct exploration of such  microscopical time scales could be of primary interest in understanding the inner nature of the quantum world -- the de Broglie ``periodic phenomenon'' has been addressed as ``the missing link'', \cite{1996FoPhL}. 
 
 BCs have  played a fundamental role since the earliest days of QM. Indeed, another advantage to using BCs as quantization conditions is  the remarkable property that QM is obtained without involving any (local) hidden variable. Since Bell's theorem is based on the fundamental hypothesis of local hidden variables, it can not be applied to our theory. On the other hand the assumption of intrinsic periodicity introduces an element of non-locality, which is however consistent with SR since the periodicity is modulated relativistically. This could be used for a novel interpretation of quantum phenomena such as  entanglement.  Thus the theory can in principle violate Bell's inequality (in particular, as we will explicitly check in future studies, if we try to adapt Bell's theorem to our theory, \ie to evaluate the expectation values of an observable in the Hilbert space described above, we expect to find again a formal parallelism with ordinary QM). For this reason we speak about---mathematical---determinism \cite{Dolce:2009ce,'tHooft:2001ar}.   \emph{We conclude that the manifestation of our description of elementary particles in terms of compact 4D and PBCs is a remarkable correspondence with relativistic QM}.

\section{\emph{Virtual} Extra Dimension (VXD) \label{sec:Virtual-Extra-Dimension}}

A de Broglie ``periodic phenomenon'' with topology $\mathbb{S}^{1}$ can be equivalently described by either assuming  intrinsic time periodicity $T_{t}(\bar{\mathbf{p}})$ in a generic reference frame as in  \cite{Dolce:2009ce}, or  by assuming intrinsic periodicity of the proper-time $T_{\tau}$ and using Lorentz transformations as shown in  sec.(\ref{Covariant:Notation}).
In this section we want to show that, from a formal point of view, the same purely 4D field theory in compact 4D described so far can be equivalently  derived from a KK theory with flat cyclic XD. This approach,  already described in \cite{Dolce:2009ce} and repeated in this section, will show us that  our pure 4D theory is manifestly \emph{dual} to a corresponding XD theory. As we will discuss in more in detail at the end of this section, in this way our description of ``periodic phenomenon'' will inherit, and for some aspects justify, interesting properties of XD theories, without actually introducing any (so far unobserved experimentally) XD.

From a mathematical point of view this can be seen by using the cyclic XD (relativistic invariant) to parameterize the intrinsic proper-time periodicity of the de Broglie ``internal clock''. That is,  the cyclic XD  will be used as a ``mathematical trick'' to describe the undulatory behavior of elementary particles. For this reason we will say that field theory in compact 4D has a ``virtual'' XD (VXD), to distinguish it from the ``real'' XD of the KK theory. 

To introduce the idea of VXD 
we consider a non-interacting 5D massless
field with ``real''   XD  denoted by $s$ in the flat metric  
\begin{equation}
dS^{2}=dx_{\mu}dx^{\mu}-ds^{2}\equiv0\,.\label{eq:5D:flat:metric}
\end{equation}
Since  an XD field with zero 5D mass is constrained on the 5D light-cone ($dS^{2}\equiv0$),  the modulo of the 4D components  and the XD component are proportional
\begin{equation}
ds^{2}=dx^{\mu}dx_{\mu}\,.\label{eq:mink:metric}\end{equation}
As already described in detail in \cite{Dolce:2009ce} and summarized in par.(\ref{Compact:space:time}),  we will identify the XD  with the world-line parameter of a purely 4D field theory. That is, we will impose  \begin{equation}
s\equiv c\tau~,\label{proptime:realtime}\end{equation}
where $\tau$ is the proper-time. To address the identification (\ref{proptime:realtime}) 
we say that the XD, $s$, is  \emph{virtual}, \cite{Dolce:2009ce}. 
  Upon this identification  the original XD theory will be reduced to a purely 4D theory with ordinary Minkowskian metric (\ref{eq:mink:metric}) (we do not consider the case of 5D massive fields since  $dS^{2}\neq0$ and this would imply tachyonic modes in the purely 4D theory).

Moreover, as in the ordinary KK theory, we will also assume that the XD is compact  with compactification length $\lambda_s$ and PBCs, \emph{i.e.} cyclic XD, \cite{Klein:1926tv}.   
Thus, with the identification (\ref{proptime:realtime}), the cyclic XD will describe a cyclic proper-time.  From this our description of a de Broglie ``periodic phenomenon'' can be derived as  in sec.(\ref{Covariant:Notation}) through the Lorentz transformation (1).
In other words,  
the original  5D theory will be reduced to a purely 4D theory;  
the KK modes of the original 5D field, which in the ``purely'' XD theory are independent 4D fields, will correspond to the harmonic modes $\Phi_{n}(x)$ of the 4D periodic field $\Phi(x)$ described in (\ref{eq:4d:periodic:field}).

\subsection{Kaluza-Klein picture}

Here we give  a short review of the formalism of the ordinary KK theory.  
A purely 5D scalar field with zero 5D mass ($dS^{2}\equiv0$) in a flat XD (\ref{eq:5D:flat:metric}) denoted by $s$, and compactification length  $\lambda_{s}$, is described by the 5D action \begin{eqnarray}
\mathcal{S}^{\lambda_{s}} & = & \frac{1}{2}\int_{0}^{\lambda_{s}}\frac{ds}{\lambda_{s}}\int d^{4}x\left[\partial_{M}\Phi^{*}(x,s)\partial^{M}\Phi(x,s)\right]\label{5D:massless}\,.\end{eqnarray}
By varying this action with respect to the 5D field $\Phi(x,s)$ and integrating by parts, we obtain 
 \begin{eqnarray}
\delta\mathcal{S}^{\lambda_{s}}  =  \frac{1}{2\lambda_{s}}\int d^{4}x\biggl\{\bigl[\delta\Phi(x,s)\partial_{5}\Phi(x,s)\bigr]_{s=0}^{s=\lambda_{s}}+ 
\int_{0}^{T_{s}}ds\delta\Phi(x,s)\partial_{M}\partial^{M}\Phi(x,s)\biggr\}.
\label{5Dmassless:variation}
\end{eqnarray}

The  bulk term of this variation yields the 5D
EoMs 
\begin{equation}
\partial_{M}\partial^{M}\Phi(x,s)=0\quad,\quad\forall s\in(0,\lambda_{s})\,.\label{eq:5deom}
\end{equation}
The generic solution $\Phi(x,s)$, through a discrete Fourier transform, can be written
as a sum over 4D normal modes $\Phi_{n}(x)$,  
\begin{equation}
\Phi(x,s)=\sum_{n}\Phi_{n}(x,s)=\sum_{n}e^{\frac{i c}{\hbar} M_{n}s}\Phi_{n}(x)\,.\label{eq:vxd:decomp}
\end{equation}
 Through the EoMs (\ref{eq:5deom}), \ie in an \emph{indirect} way, these can be interpreted as 4D massive fields, \aka the KK mass eigenmodes. In fact 
\begin{equation}
\left(\partial_{\mu}\partial^{\mu}+\frac{M_{n}^{2} c^{2}}{\hbar^{2}}\right)\Phi_{n}(x,s)=0\,.\label{eq:vxd:decomp:bulkEoM}
\end{equation}

The fundamental requirement is, as usual, that the BCs must minimize the
action (\ref{5D:massless}) at the boundary, \emph{i.e.}  that
the boundary term of (\ref{5Dmassless:variation}) must vanish.
In the KK theory,  this requirement  is fulfilled by assuming PBCs (though  anti-PBCs or, more in general, N and D BCs are equally allowed). The resulting  theory with cyclic XD has topology $\mathbb S^{1}$. 
We may already note that the  PBCs at the ends of the XD 
\begin{equation}
\Phi(x,s)\equiv\Phi(x,s+\lambda_{s})\,,\label{eq:prop:time:period}
\end{equation}
lead \emph{indirectly} (see comments at the end of this section) to a KK mass tower  $M_{n}$. This turns out to be formally the same mass spectrum of field theory in compact 4D given in (\ref{eq:KK:energy:spectr}), \ie $M_n = n \bar M = n h / \lambda_s c$.

Integrating over the XD, the KK action (\ref{5D:massless}) can
be equivalently written as a sum ($n\in\mathbb{Z}$) over the 4D Lagrangian
densities of the KK mass modes $\Phi_{n}(x)$:  
\begin{equation}
{\mathcal{S}}^{\lambda_{s}}=\sum_{n}\int d^{4}x\frac{1}{2}\Bigl[\partial_{\mu}\Phi_{n}^{*}(x)\partial^{\mu}\Phi_{n}(x)-n^{2}\frac{{\bar M}^{2} c^{2}}{\hbar^{2}} \Phi_{n}^{2}(x)\Bigr]\,.\label{eq:action:VKK}
\end{equation}

Next we will show the dualism between field theory compact 4D and the  KK theory described here, by explicitly assuming that the XD is \emph{virtual},  (\ref{proptime:realtime}). The 4D modes $\Phi_{n}(x)$ which in the KK theory are independent fields will be the energy excitations of the same fundamental 4D field  (\ref{eq:4d:periodic:field}). 

\subsection{Dualism to field theory in flat XD}\label{Compact:space:time}

Here we explicitly assume that the cyclic XD of the KK theory introduced above is a \emph{virtual} XD rather than a ``real'' XD.  
This means to impose the condition (\ref{proptime:realtime}). The resulting purely 4D theory with cyclic world-line parameter will be nothing other than the field theory in compact 4D with Minkowskian metric (\ref{eq:mink:metric}) described in sec.(\ref{Covariant:Notation},\ref{sec:Correspondence-with-QFT}). That is,  $\Phi(x,s)$ in (\ref{eq:vxd:decomp}) will describe the same de Broglie ``periodic phenomenon'' as the purely 4-periodic field $\Phi(x)$.  

The condition of VXD, (\ref{proptime:realtime}), means that the parameter $s$ must be identified with a cyclic world-line parameter
(implicit function of the 4-coordinate $x_\mu=\{c t,- \mathbf{x}\}$)   with periodicity $\lambda_{s}$.  This corresponds to imposing that the proper-time $\tau$ has intrinsic periodicity 
$T_{\tau}={\lambda_{s}}/{c}$, according to (\ref{Compton:mass:period}). 
From this it is easy to see that a bosonic field theory with such a proper-time periodicity  is nothing other than our field theory in compact 4D. By following the line of par.(\ref{Covariant:Notation}), it is sufficient to note that,  if we move a de Broglie ``periodic phenomenon'' of intrinsic proper-time periodicity $T_{\tau}$ from its rest frame to a generic frame denoted by the spatial momentum $\mathbf{\bar{p}}$,
we get a resulting time periodicity $T_{t}(\mathbf{\bar{p}})$, \cite{Broglie:1924,1996FoPhL}. That is, in a generic reference frame, 
according to the Lorentz transformation (\ref{Lorentz:prop:time}),  the proper-time periodicity (\ref{eq:prop:time:period}) induces the following time periodicity 
\begin{equation}
\Phi(\mathbf{x},t,s)\equiv\Phi(\mathbf{x},t+T_{t}(\mathbf{\bar{p}}),s)\,.\label{eq:real:time:period}
\end{equation}

From this  we can for instance repeat the same demonstration used in \cite{Dolce:2009ce} obtaining back field theory in compact 4D of sec.(\ref{Covariant:Notation}). 
In fact, the intrinsic  time periodicity (\ref{eq:real:time:period}) implies that the 4D field $\Phi(x,s)$ in  (\ref{eq:vxd:decomp})
 can be further
expanded as  three-dimensional energy eigenmodes $\Phi_{n}(\mathbf{x})$ 
\begin{equation}
\Phi(x,s)  =  \sum_{n}e^{-\frac{i n}{\hbar}(\bar{E}(\mathbf{\bar{p}})t-\bar{M}c s)}\Phi_{n}(\mathbf{x})\,.
\end{equation}
The resulting quantization of the energy spectrum associated to the time periodicity $T_{t}(\mathbf{\bar{p}})$ of the field (\ref{eq:real:time:period}) is the same harmonic energy spectrum (\ref{eq:energyspectr:period}) of field theory in compact 4D, \emph{i.e.} $E_n (\mathbf{\bar{p}}) = n \bar E (\mathbf{\bar{p}}) = n h / T_{t}(\mathbf{\bar{p}})$.
In particular, the expansion in energy and mass eigenmodes of $\Phi(x,s)$ is labeled  by the same (quantum) number  $n$ because the corresponding time or world-line periodicities are relativistic projections of the same de Broglie ``clock'' of fundamental topology $\mathbb S^{1}$. 
Note that  in the rest frame ($\mathbf{\bar{p}}\equiv 0$) we have the matching condition  
$T_{t}(0)=T_{\tau}$.

After decompactification  of the time coordinate, the action (\ref{eq:action:VKK}) turns out to be   a single sum over 3D Lagrangians of the energy eigenmodes $\Phi_{n}(\mathbf{x})$,
 \begin{equation}
{\mathcal{S}}^{\lambda_{s}}=T_{t}\int^{\mathbb{R}^{3}}d^{3}x\sum_{n} \frac{1}{2}\left[\frac{n^{2}}{c^{2}\hbar}\left( {\bar{E}^{2}(\bar{\mathbf{p}})^{2}} - {\bar M^{2} c^{4}}\right)\Phi_{n}^{2}(\mathbf{x})-|\partial_{i}\Phi_{n}(\mathbf{x})|^{2}\right]\,.\label{proper-real-time:period:action}
\end{equation}

In order to find the dispersion relation of the 4D eigenmodes $\Phi_{n}(x)$ we must resolve the EoMs of (\ref{proper-real-time:period:action}), so that
$n^{2}{\bar{E}^{2}(\bar{\mathbf{p}})}=\mathbf{p}_{n}^{2} c^{2}+n^{2}{\bar{M}^{2}}{c^{4}}$. 
From this it is easy to see that  these eigenmodes are described in a collective or coherent way and that  the quantization of the momentum spectrum is  $\mathbf{p}_{n} = n \mathbf{p}$. 
In fact, as noticed for instance with the Lorentz transformation (\ref{Lorentz:prop:time}), in a generic reference frame the intrinsic proper-time periodicity (\ref{eq:real:time:period}), or equivalently the periodicity of the VXD  (\ref{eq:prop:time:period}), together with the time periodicity $T_{t}(\mathbf{\bar{p}})$, also induces a periodicity $\lambda_{x}$ (wave-length) on the modulo of the spatial dimensions.  The quantization of the momentum spectrum in terms of this spatial periodicity is given by (\ref{eq:quant.wavenum:def}).
In the rest frame (\textit{i.e.} $d\tau\equiv dt$) we have $\mathbf{p}_{n}=0$,  $\forall n$ (\textit{i.e.} zero separation $d\mathbf{x}\equiv0$) and the matching condition   
$\bar{E}(0)=\bar{M} c^{2}$. 
Thus, the fundamental mode has dispersion relation of a relativistic particle (\ref{relat:constr:4p}), and the resulting space-time periodicity $T^\mu$ is the usual de Broglie 4-periodicity  (\ref{Lorentz:transf:T})  with geometrical constraint (\ref{relat:constr:4T}). From this follows that  the resulting dispersion relation of the quantized energy spectrum is again that of a ``periodic phenomenon'' (\ref{eq:normal:ordered:spectr}), which in turn matches with the energy spectrum  of a second quantized KG field with mass $\bar M$ (after normal ordering).

The action (\ref{5D:massless}) or (\ref{eq:action:VKK}) can be regarded as  purely 4D under the assumption of VXD. In fact, in (\ref{eq:action:VKK})   the explicit dependence on the world-line parameter of  (\ref{5D:massless}), \emph{i.e.} the VXD, has been integrated out.
Since the whole information of the decompactified action (\ref{eq:action:VKK}) is contained in a single space-time period, we can limit the integration region of the action to  $T^{\mu}$ --- analogous arguments apply to the spatial integration region of (\ref{proper-real-time:period:action}). 

In conclusion, the field solution $\Phi(x,s)$ of a KK theory with VXD  is actually the periodic solution $\Phi(x)$, (\ref{eq:cond:4D:period}), of field theory in compact 4D (\ref{generic:actin:comp4D:transf}). The KK modes, under the assumption of VXD, turn out to be the energy excitations of the de Broglie ``periodic phenomenon''. They can be named \emph{virtual} KK modes, for reasons that will be described in the comments below.

To pass from the VXD field $\Phi(x,s)$ satisfying
(\ref{5D:massless}) and (\ref{eq:prop:time:period}) to the corresponding 
4D field $\Phi(x)$, (\ref{eq:4d:periodic:field}), satisfying (\ref{generic:actin:comp4D}) we integrate out the explicit dependence on the world-line parameter of the field as follows: 
\begin{equation}
\Phi(x)=\sum_{n'}\int_{0}^{\lambda_{s}}\frac{ds}{\lambda_{s}}e^{+i{M}_{n'} c s}\Phi(x,s)\,.\label{VXDto4D:field}
\end{equation}

Notice that, as long as  PBCs are assumed and $s$ is identified with the world-line parameter, (\ref{5D:massless}), (\ref{eq:action:VKK}) and (\ref{proper-real-time:period:action}) are equivalent formulations of the same purely 4D action (\ref{generic:actin:comp4D}) describing our de Broglie ``periodic phenomenon''. 
By imposing the bulk EoMs (\ref{eq:5deom}) as a constraint, the action in a VXD (\ref{5D:massless}) can also be rewritten as  a 4D boundary action \begin{equation}
{\mathcal{S}}^{\lambda_{s}}=\frac{{1}}{2}\int^{T_{t}(\bar{\mathbf{p}})}dt\int^{\mathbb{R}^{3}}d^{3}x\left[\Phi(x,s)\partial_{5}\Phi(x,s)\right]_{s=0}^{s=\lambda_{s}}~.\label{eq:act:virt:hologr}\end{equation}
In this formulation we say that  the bulk physics of the VXD  has been projected on the boundaries.  This reduction is at the base of the holographic prescription that we will discuss later. In (\ref{eq:act:virt:hologr})  we have not yet explicitly imposed any BCs at the ends of the VXD.
 If we impose PBCs  and the bulk EoM as constraints, (\ref{eq:act:virt:hologr}) describes the same physics of (\ref{generic:actin:comp4D}) as well.

 \subsection{Comments and  Outlooks}

Field theory in XD has  remarkable mathematical properties such as  Kaluza's unification of gravity and electromagnetism \cite{Kaluza:1921tu}; an elegant explanation for the hierarchy of about 17 orders of magnitude between the electroweak scale and the Planck scale \cite{ArkaniHamed:1998rs}; and more recently, with the warped models, interesting interpretations of the fermion mass hierarchy \cite{Gherghetta:2006ha}, of the electroweak gauge symmetry and unitarization of the elastic scattering without involving any additional fundamental scalar (\eg Higgsless \cite{Csaki:2003dt}, Gauge-Higgs-Unification or composite-Higgs models).  
On the other hand, no experimental evidence of the existence of an XD has been obtained so far. This has strongly constrained some of the above XD theories, which in turn are sometimes regarded as of purely mathematical interest.  

The idea emerging from our analysis of KK theory, and its reduction to a purely 4D theory in compact dimensions, is that some of the good properties of XD theories can be justified without actually introducing any ``real'' (unobserved) XD. This is because, as we have seen, the role played by  the proper-time  of a de Broglie ``periodic phenomena'' is, from a mathematical point of view,  exactly analogous to the one played by an XD in the KK theory.   

We want now give conceptual elements to interpret the dualism between a free field in compact 4D and a 5D field in flat compact XD formally obtained  above. To see this we must consider that, in sec.(\ref{Covariant:Notation}), according to undulatory mechanics,  we have described  an isolated elementary particle as an intrinsically periodic phenomenon in which the proper-time can be regarded as cyclic. In this section we have seen that the same undulatory description can be retrieved from a simple 5D field in flat cyclic XD simply by identifying the  XD as the cyclic proper-time.  Applying this recipe, it is easy to figure out that \emph{the resulting  theory inherits fundamental mathematical properties of the initial XD theory, though remaining a purely 4D theory}.

In doing this identification  
we say that we impose that the XD is \emph{virtual}, or equivalently that the cyclic world-line parameter plays the role of a \emph{virtual} XD. The term \emph{virtual} XD is loosely used to highlight the fact that in either case the resulting field theory in compact 4D is purely 4D, though it shares some good behavior with the XD theory.    
Another reason for using the term \emph{virtual} has been already mentioned during the conceptual interpretation of the quantum behavior associated to a ``periodic phenomenon''.  In fact,  the energy excitations of a periodic field can be regarded as quantum excitations or, roughly speaking, as \emph{virtual} particles,  according to the  correspondence to relativistic QM described in the previous section. But we have also seen here that,  under the assumption of VXD, these energy excitations correspond to the KK modes of the original XD, so that they can be addressed as \emph{virtual} KK modes. 
 We will come back to this point when we will discuss about AdS/CFT and hadrons.

 Note that in an XD  theory the mass spectrum arises \emph{indirectly}  in (\ref{eq:vxd:decomp:bulkEoM})  through the  EoMs (\ref{eq:5deom}). That is, the quantized conjugate variable of 
the XD can be identified with a mass spectrum because $d S^2 \equiv 0$  (this is analogous to the fact that, through the EoMs, the modulo of the spatial momentum is proportional to the energy in massless particles). On the other hand,  in the case of field theory in compact 4D, the mass spectrum arises \emph{directly} through a discrete Fourier transform: it is the quantized ``physical  conjugate'' variable of the compact world-line parameter. Thus it can be named a \emph{virtual} KK tower. 
In the ordinary XD theory  the KK modes are independent 4D  fields (with unrelated 4-momenta)  whereas under the assumption of VXD they turn out to be the excitations of the same 4D field. Indeed the  \textit{virtual} KK tower of the 4D field $\Phi(x)$ is a \emph{coherent} sum  of energy  eigenmodes $\Phi_n(x)$, \ie the  \textit{virtual} KK modes  have a \emph{collective} description since they all satisfy the same 4D PBCs. Such a \emph{collective} description of the  energy eigenmodes (\ref{eq:4d:periodic:field}) as \emph{virtual} KK modes shows also an interesting aspect of the analogy of field theory in compact 4D with the Matsubara theory. As is well known, statistical systems are quantized by using the ``mathematical trick''  of an intrinsic Euclidean time periodicity, so that the 4D  Euclidean field solution can be decomposed into a \emph{coherent} sum of thermal energy eigenmodes.
 We will see that a \emph{collective} description of the KK modes typical of a VXD is implicit in the usual holographic description of an ordinary KK theory. 
The compactification length of ordinary XD theories is a free parameter, which however must be extremely small since no XD has been  experimentally observed so far. Current theory sets this value to be smaller than $10^{-19} m$, \ie to an energy scale above TeV. With the assumption of VXD, every elementary system has associated a different compactification length, uniquely fixed by the inverse of the mass. This assumption is completely falsifiable, and according to the previous section has remarkable correspondences to ordinary relativistic QM. We will see later that our theory also has fundamental analogies with string theory. In this way it will be  natural to interpret the hadrons as quantum excitations of the same elementary string.

\section{Interactions\label{sec:Towards-Interactions}}

An exhaustive formalization of interacting  periodic fields,
and thus the exact transition between  different periodic regimes, is beyond the scope of this paper. 
Here we want to give a qualitative description
by considering the deformations of the boundary induced by  particularly  simple interaction schemes. A more rigorous  formalization of the problem  is given in \cite{Dolce:tune}.  From this section on, we adopt natural units ($\hbar\equiv c \equiv1$).

A simple starting point to introduce our formalization of interactions is to consider  the Compton scattering 
$e+\gamma\rightarrow e'+\gamma'$.
 According to the de Broglie  (\ref{eq:4D:debroglie:period}),  the conservation of the 4-momentum 
$\bar{p}_{\mu}^{e}+\bar{p}_{\mu}^{\gamma}=\bar{p}_{\mu}^{e'}+\bar{p}_{\mu}^{\gamma'}$ 
in the interaction point $x=X$, 
can be equivalently expressed in terms of conservation
of the inverse of the 4-periodicity, 
$({T_{e}^{\mu}})^{-1}+({T_{\gamma}^{\mu}})^{-1}=({T_{e'}^{\mu}})^{-1}+({T_{\gamma'}^{\mu}})^{-1}$.

With this example we see that local and retarded variations of the 4-momentum
occurring in the interaction point can be equivalently interpreted
as  local and retarded modulations of the 4-periodicity of the fields. In the formalism of field theory in compact 4D, modulation of periodicity corresponds to
 a local stretching of the compactification 4-length, and thus to a local deformation of the metric. Therefore, to describe the interaction we will use the formalism field theory in curved 4D \cite{Birrell:1982ix}. 
  In particular, as is well know from the geometrodynamical description of gravitational interaction of GR, the local modulations of periodicity of reference clocks can be equivalently encoded in local deformations of the underlying space-time metric.

\subsection{Geometrodynamics of the boundary}\label{Geometrodynamics:boundary}

In classical-relativistic mechanics a particular interaction scheme can be described in terms of corresponding local variations of 4-momentum with respect to the non-interacting case. 
That it, 
in a generic interaction point $x=X$, it can be described by the following local variation of 4-momentum 
\begin{equation}
\bar{p}_{\mu}\rightarrow\bar{p}'_{\mu}(X)=e_{\mu}^{a}(x)|_{x=X}\bar{p}_{a}\,.\label{eq:deform:4mom:generic:int}
\end{equation}
With this notation we mean that, as the interaction is switched on,  the persistent 
4-momentum $\bar{p}_{a}$ of a free elementary system is forced  to vary from point to point.  
The specific interaction scheme  is therefore encoded  in the tetrad $e_{\mu}^{a}(x)$. We now generalize the analysis given for the Lorentz invariance (\ref{Lorentz:transf}) to the more general case of invariance of the theory under local  transformations of variables, see also \cite{Dolce:tune}.

Because of the relation between the 4-momentum and the 
 compactification 4-length  (\ref{eq:4D:debroglie:period}), the interaction (\ref{eq:deform:4mom:generic:int}) can be equivalently encoded in a corresponding local contravariant modulation of 4-periodicity with respect to the persistent periodic case:
\begin{equation}
T^{\mu}\rightarrow T'^{\mu}(X)\sim e_{a}^{\mu}(x)|_{x=X}T^{a}\,.\label{eq:deform:4period:generic:int}
\end{equation}
In the approximations of this paper  the 4-periodicity can be identified  with the  compactification 4-length of the theory \cite{Dolce:tune}.
Thus,  during interaction we have a corresponding  
stretching of the compact 4D and in turn a deformation of the flat 4D metric,  
\begin{equation}
\eta_{\mu\nu}\rightarrow g_{\mu\nu}(x)=e_{\mu}^{a}(x)e_{\nu}^{b}(x)\eta_{ab}\,.\label{eq:deform:metric:generic:int}
\end{equation}
To check  this geometrodynamical description it is sufficient to consider the  transformation of variables associated with the interaction scheme (\ref{eq:deform:4mom:generic:int}-\ref{eq:deform:metric:generic:int}),  
\begin{equation}
dx_{\mu}\rightarrow dx'_{\mu}(X)=e_{\mu}^{a}(x)|_{x=X}dx_{a}\,,\label{eq:deform:mesure:generic:int}
\end{equation}
as a substitution of variables in the free action (\ref{generic:actin:comp4D}). The determinant of the Jacobian is  $\sqrt{-g(x)}=\det[e^{a}_{\mu}(x)]$. In this way we actually obtain \cite{Birrell:1982ix} 
\begin{equation}
\mathcal{S}^{\lambda_{s}}\sim\oint_{0}^{e_{a}^{\mu}(X)T^{a}}\!\!\!\!\!\!\!\!d^{4}x\sqrt{-g(x)}\mathcal{L}(e^{\mu}_{a}(x)\partial_{\mu}\Phi'(x),\Phi'(x))\,.\label{eq:defom:action:generic:int}\end{equation}
 This transformed action has  compactification 4-length $T'^{\mu}(X)$ varying locally. 
 Hence, its field solution has modulated periodicities at different interaction points $X$. This actually describes the modulation of de Broglie periodicity of a particle under the interaction scheme (\ref{eq:deform:4mom:generic:int}).  
Note that the geometrical description given in  this paper 
 is limited  to the cases in which the  compactification 4-length of the theory can be approximated with the local 4-periodicity; for a more general formalism see  \cite{Dolce:tune}.

According to our analysis,  interactions can be formalized in terms of local
diffeomorphisms (\ref{eq:deform:mesure:generic:int}), 
\emph{i.e.} in terms of  field theory in curved 4D (\ref{eq:defom:action:generic:int}). 
The free case is given by the classical field 
solution with persistent periodicity $T^\mu$  associated to 
 the  bosonic action in flat 4D and persistent boundary
(\ref{generic:actin:comp4D}).
The interaction
scheme (\ref{eq:deform:4mom:generic:int})  is described
by 4-periodicity  local modulations  of the de Broglie ``internal clocks'' of the particles and encoded in the corresponding deformed metric (\ref{eq:deform:metric:generic:int}).  As we will discuss below, this mimics  GR. 
In fact, the gravitational interaction can be interpreted as modulations of spatial and temporal periodicity of reference lengths and clocks, respectively, encoded in  corresponding deformed space-time metric.

The conditions at the transformed boundary $T'^{\mu}$ must be of the same
type as the ones assumed for the free fields of (\ref{generic:actin:comp4D}) at the persistent boundary $T^{\mu}$.
In our case, to reproduce the quantum behavior, we  assume PBCs
(or conditions describing the same physics at an effective level, as in the holographic formulation).
In this case the matching between the fundamental mode of a cyclic field and the corresponding KG mode is local, \cite{Dolce:tune}.

To extend this geometrodynamical description of interactions to the VXD formalism, the flat VXD  metric describing a free cyclic field must be deformed in the following way 
\begin{equation}
\eta_{M N}\rightarrow g_{M N}(s)=e_{M}^{A}(s)e_{N}^{B}(s)\eta_{AB}\,.
\end{equation}
For the scope of this paper  it is sufficient to assume only  deformations of the
4D components of the \emph{virtual} metric
\begin{equation}
g_{M N}(s)\sim\left(\begin{array}{cc}
g_{\mu\nu}(x(s)) & 0\\
0 & 1\end{array}\right)\,,\label{VXD:deform:metric}
\end{equation}
where the capital  letters label the 5D
Lorentz indices. In fact, in this paper, such a formalism will be exclusively used  to describe massless fields (a  consistent generalization to massive fields may involve, for instance, dilatons in the deformed VXD metric). Thus, to describe interactions of massless fields in the VXD formalism, (\ref{5D:massless}) must be transformed to an action in the corresponding curved 5D (\ref{VXD:deform:metric}). The resulting  theory in curved 4D  (\ref{eq:defom:action:generic:int}) is obtained by imposing that the XD is \emph{virtual}, (\ref{proptime:realtime}), \ie by integrating out the VXD as prescribed in sec.(\ref{sec:Virtual-Extra-Dimension}). 

\subsection{Matching with relativistic geometrodynamics}\label{relativistic:geometrodynamics}

As already noticed for instance in \cite{Dolce:tune}, the analogy with GR arises naturally if we think of the de Broglie internal clocks of elementary particles as reference clocks. In fact gravitational interaction  can be described as modulations of 4-periodicity and thus encoded in corresponding local deformations of the underlying 4D metric.

More explicitly 
we consider the  case of weak Newtonian interaction instead
of the generic interaction (\ref{eq:deform:4mom:generic:int}).
In a weak  gravitational potential 
\begin{equation}
V(\mathbf{x})=-\frac{GM_{\odot}}{|\mathbf{x}|}\ll 1\,,\label{eq:low:grav:pot}
\end{equation}
according to (\ref{eq:deform:4mom:generic:int}), the variation of the energy with respect to the non-interacting case 
is \cite{Ohanian:1995uu} \begin{equation}
\bar{E}\rightarrow\bar{E}'\sim\left(1+\frac{{GM_{\odot}}}{|\mathbf{x}| }\right)\bar{E}\,.
\end{equation}
Thus, according to (\ref{eq:deform:4period:generic:int}) and to the de Broglie phase harmony (\ref{eq:4D:debroglie:period}), the corresponding
modulation of  time periodicity is 
\begin{equation}
T_{t}\rightarrow T_{t}'\sim\left(1-\frac{{GM_{\odot}}}{|\mathbf{x}|}\right)T_{t}\,.
\end{equation}
Indeed, see \cite{Ohanian:1995uu}, this simple description  is already sufficient to retrieve two important predictions of GR: gravitational redshift and the usual gravitational dilatation
of  the Minkowskian time (as arises explicitly from the  Hafele and Keating experiment this aspect is described in terms of variation of the periodicity of reference clocks). Essentially, this is
a direct consequence of the modulation with the energy  of the de Broglie internal clocks, which therefore run slower in a gravitational well with respect to the non-interacting case, according to  (\ref{eq:4D:debroglie:period}).
Thus, our formalism mimics linearized gravity
 
$g_{\mu\nu}(x)\simeq\eta_{\mu\nu}+\kappa h_{\mu\nu}(x)$. 
 In fact, from (\ref{eq:deform:metric:generic:int}) we explicitly obtain
the contribution $h_{00}= -\frac{1}{\kappa}\frac{GM}{|\mathbf{x}|}$ to the deformed metric.
Besides the energy variation and time modulation we must consider the variation of the
spatial momentum (\ref{eq:deform:4mom:generic:int})  induced by the gravitational interaction and the consequent modulation of the spatial periodicity (\ref{eq:deform:4period:generic:int}). In this way it is possible to find out that (\ref{eq:deform:metric:generic:int}) is nothing other than 
the usual Schwarzschild metric in the linear approximation
(\ref{eq:low:grav:pot}), see for instance \cite{Ohanian:1995uu}. Furthermore, it is well known that GR can be derived from the linearized formulation by considering self-interactions \cite{Ohanian:1995uu,Dolce:Dice} --- for instance by relaxing the assumption of smooth interactions.
 
\subsection*{Comments and Outlooks} 
 
In this section we have shown that the geometrodynamics encoding interactions in the compact 4D formalism mimic the relativistic geometrodynamics of linear gravity. 
Though the exact formalism must be explicitly worked out and checked, this provides  evidence 
 of the consistency of our  formalism of  compact 4D
with GR.

To understand this description it is important to point out that ``\emph{what is fixed at the boundary of the action principle of GR}''  is not uniquely defined \cite{springerlink:10.1007/BF01889475}. SR and GR fix the differential framework of the 4D without giving any particular prescription about the BCs. The only requirement for the BCs is that to minimize a relativistic action at the boundary. For this aspect both SBCs and PBCs have the same formal validity and consistency with relativity. Thus \emph{SR and GR allow us the freedom to play with BCs}, as long as we fulfill the variational principle; and the BCs have played a fundamental role since the first days of QM, according to de Broglie, Planck, Sommerfeld, Bohr, Klein, etc. Nevertheless, the BCs play a very marginal role in ordinary QFT. This has motivated the implementation of our boundary methods in field theory, showing at the same time the possibility that \emph{quantum phenomena can be interpreted semi-classically by playing with  BCs in relativistic theory}. These considerations  may also offer novel elements to address the problem of the quantization of gravity.    
 
 It is interesting to point out that, from a historical point of view, Einstein formulated  GR by thinking of the modulations of periodicity of reference clocks in a gravitational potential. 
 Indeed, in general relativity, the local deformation of the space-time metric encodes the modulations of periodicity of reference clocks.  In perfect analogy with this description, the formalism of field theory in compact 4D,  associated to every point a local 4-vector $T^\mu(X)$ describing the modulations of periodicity of the internal clocks of interacting elementary particles, which in turn can be equivalently encoded to deformations of the underlying flat metric to locally curved metrics (\ref{eq:deform:metric:generic:int}). The local modulations of de Broglie 4-periodicity describe the local variations of 4-momentum associated to the interaction scheme (\ref{eq:deform:4mom:generic:int}).  This also means that in this formulation the kinematics of the interaction  is encoded on the relativistic geometrodynamics of the boundary of the theory (\ref{eq:defom:action:generic:int}). Hence we have an interesting analogy with the holographic principle \cite{'tHooft:1993gx,Susskind:1994vu}.

 To figure out the conceptual consistency of our description, we must remember the considerations given in sec.(\ref{Sec:Compact:spacetime}), when we noticed that massless fields have infinite world-line compactification length (frozen proper ``internal clock''), so that they can be used as reference for a relational description of the massive particles constituting the system under investigation. This description can be extended to interactions as the modulation of periodicity is considered. In particular, gravitational interaction, as EM interaction, being mediated by massless fields, naturally sets the reference space-time coordinates to describe the reciprocal dynamics of the gravitating particles. That is, the gravitational field can have infinite proper-time periodicity, so that a system of particles interacting gravitationally can be described, as in usual non-cyclic cosmological theory,  by a non-compact world-line parameter.

Remarkably, in \cite{Dolce:tune} we have argued that the same approach can be used to achieve a geometrodynamical description of gauge interactions.  
In a few words, the gauge field turns out to encode the modulation of periodicity associated to local transformations of flat reference frames. The resulting classical evolution of the modulated de Broglie ``periodic phenomenon'', with all its harmonic modes, turns out to be described by the ordinary FPI  scalar QED, \cite{Dolce:tune}.  

\section{AdS/CFT and Holography}\label{sec:AdSCFT:interpr}

In this section we review basis aspects of AdS/CFT in field theory.
In modern physics, 
one of the most important issues which seems to relate classical geometrodynamics 
with  quantum dynamics is the AdS/CFT correspondence.
Conjectured by Maldacena in 1997 \cite{Maldacena:1997re} the original
formulation states that, in the 't Hooft limit $g_{YM}^{2}N_{c}\gg1$,
the IIB string theory on $AdS_{5}\times\mathbb{S}^{5}$ is dual to
the $\mathcal{N}=4$ supersymmetric $SU(N)$ Yang-Mills theory, where
$g_{YM}$ is the Yang-Mills coupling.
Roughly speaking a \emph{tree level} string theory on an Anti-de Sitter (AdS) background turns out to be dual to a corresponding Conformal Field Theory (CFT) Strongly Coupled (SCFT), \cite{Gherghetta:2006ha}.

As summarized by  Witten in the abstract of his paper  \cite{Witten:1998zw}: under the AdS/CFT correspondence ``\emph{quantum phenomena [...] are encoded in classical geometry}'' without introducing any explicit quantization condition. Such a classical to quantum correspondence of AdS/CFT has been tested for several non-trivial aspects of modern physics, including condensed matter, and it seems to have a more general validity with respect to its original string formulation. We will investigate the semi-classical description of relativistic QM obtained so far in terms of some basic phenomenological (4D) aspects of AdS/CFT \cite{Gherghetta:2006ha,Contino:2004vy,ArkaniHamed:2000ds}. 

For our purpose we consider the simple phenomenological formulation in field theory of the AdS/CFT correspondence given
through the holographic prescription \cite{Witten:1998zw,Gherghetta:2006ha,Contino:2004vy}.  Similarly to (\ref{eq:act:virt:hologr}) and (\ref{VXD:holo:appr}),
holography consists of projecting the classical bulk dynamics on the
boundary by imposing as constraint to the field $\Phi$ the bulk
EoMs and fixed value $\Phi|_{\Sigma}=e\phi_{\Sigma}$ at the boundary $s = \Sigma$.
 In this way we are left with a lower dimensional theory where the only
relevant \emph{d.o.f.} is the so-called source or interpolating field $\phi_{\Sigma}$.
The coupling of the source field with the bulk field is given by the
parameter $e$. In this holographic prescription, a XD 
theory is effectively described  by 
the following  4D holographic action: 
\begin{equation}
{\mathcal{S}}_{\phi_{\Sigma}}^{Holo}=\mathcal{S}^{Kin}_{\phi_{\Sigma}} +
 \int\frac{d^{4}p}{8 \pi^{2}}\left[\phi_{\Sigma}(-p)\phi_{\Sigma}(p)\Pi^{Holo}(p^{2}))\right]~.\label{source:holo:act}
\end{equation}
Here we have written the action in the 4-momentum space. 
As we will see below, the holographic action  encodes in an effective way the information of the  classical bulk configurations
of the XD  fields.  By associating 
a conformal operator ${\cal O}$ with source field $\phi_{\Sigma}(p)$ to the bulk field $\Phi$
and assuming an AdS metric, the central meaning of the AdS/CFT correspondence is
 summarized by the relation \cite{Witten:1998zw}\footnote{The complete theoretical formulation of the conjecture involve a further functional integration on both sides of (\ref{AdS:CFT:corr}) over all the kinematic configurations of $\phi_{\Sigma}$. We will not consider this aspect since the theory in VXD describes the quantum behavior without any further quantization.}
 \begin{equation}
 \int\mathcal{D}\phi_{SCFT}e^{{iS^{SCFT}[\phi^{SCFT}]+ie\int dx\phi_{\Sigma}\mathcal{O}}}
 \leftrightsquigarrow
  e^{{i{\mathcal{S}}_{\phi_{\Sigma}}^{Holo}}}\,.\label{AdS:CFT:corr} \end{equation}
  
 In other words, under the AdS/CFT correspondence any given
\emph{classical} configuration of a XD field in the AdS bulk
has in the holographic formulation, a dual interpretation in terms
of a 4D  \emph{quantum} SCFT, \cite{Gherghetta:2006ha}. Thus, we have  the remarkable property that the classical
partition function of a 4D holographic theory is related
to the quantum generating functional of a SCFT \cite{Witten:1998zw,Gherghetta:2006ha}. 

Under this correspondence  the holographic correlator
$\Pi^{Holo}(p^{2})$ describing the AdS classical dynamics of the fields is dual to the
quantum two-point function $\left\langle \mathcal{O}(p)\mathcal{O}(-p)\right\rangle $
of the SCFT. The correspondence to CFT is when the holographic dimension $s$ is non-compact (infinite compactification length), whereas if we impose branes,  the conformal symmetry turns out to be broken.

 The holographic
prescription is an useful computational technique to achieve the effective
description of XD models  (not necessarily in an AdS background) avoiding the explicit summation 
over the  KK mass eigenmodes. The effective theory is obtained 
by  choosing the correct source fields and couplings to
the bulk fields. Typically, \cite{Barbieri:2003pr,Casalbuoni:2007xn}, the source field $\phi_{\Sigma}$ and the coupling $e$ can be approximated with the  fundamental KK mode $\bar \phi$ and the  boundary value  of its delocalization profile: $\bar \Phi|_{\Sigma} \sim e \phi_{\Sigma} $.
From a computational point of view, this  corresponds to  ``integrating out'' the heavy modes of
the KK tower from the EoMs, \cite{Casalbuoni:2007xn}. The poles of the holographic correlator encode the KK mass spectrum of the corresponding XD theory. Therefore, the holographic description (\ref{source:holo:act})  automatically corresponds to the low energy
effective description of the underlying XD theory, so that it can be used for the phenomenological study of XD models \cite{Gherghetta:2006ha,Contino:2004vy,Barbieri:2003pr,Casalbuoni:2007xn}. 

\subsection*{Comments and  Outlooks}

At this point a further digression about the  holography prescription is necessary.
The choice  of source field and coupling described above corresponds to BCs effectively describing the N-BCs (or similarly of the PBCs or anti-PBCs) of the underlying XD theory. These ``holographic'' BCs play a role similar to the ordinary SBCs of QFT. In both cases they fix the boundary value of the field.
In the ordinary  SBCs the boundary value of the field is not ``dynamical'' (the Fourier coefficients of the  generic field solution are constants). For instance SBCs can be used to pick up only the fundamental mode of the tower, \ie  to ``eliminate'' completely the higher modes of the theory. In the formalism of compact 4D,  SBCs (instead of PBCs) can be used to describe the non-quantum limit, eliminating all the energy eigenmodes but the fundamental one. In fact, we have already mentioned that SBCs can be used to select the KG mode associated to the fundamental mode of the periodic field solution, see \cite{Dolce:tune}.

     In the holographic prescription, the boundary value of the field, \textit{i.e.} the source field,  is ``dynamical'' (in the holographic action it has a kinetic term $\mathcal{S}^{Kin}_{\phi_{\Sigma}}$  and the Fourier coefficient of the  field solution are functions, such as Bessel functions in an AdS metric, containing poles). 
     Thus, the holographic solution  is not a simple single mode $\bar \Phi$ of the tower. This implies that the holographic correlator
$\Pi^{Holo}(p^{2})$ and its pole  encodes the effective  propagation and the spectrum of the higher KK modes, respectively.  Thus, another interesting consequence of such an holographic prescription is that the KK modes are described in a \emph{collective}  way. In fact they all depend on the single  4D \emph{d.o.f.} $\phi_{\Sigma}$. They all fulfill the same 4D BCs similarly to the case in which the XD is \emph{virtual}. It is important to point out that such a  \emph{collective} description is  similar to the one obtained in the VXD formulation where the KK modes are \emph{virtual}, \ie they are the \emph{collective} modes of the same 4D \emph{d.o.f.}, and can be interpreted as the  excitations of the same quantum system.

\section{XD classical geometry to 4D quantum behavior correspondence \label{sub:VXD/QFT--correspondence}}

By combining together all the  results obtained so far we will see that, through
the assumption of PBCs, the classical geometrodynamics of the deformed 4D boundary of (\ref{eq:defom:action:generic:int}) 
 turns out to encode the quantum behavior of the corresponding interaction
scheme. This aspect of the theory has been recently used in \cite{Dolce:tune} to realize a geometrodynamical interpretation of scalar QED. In this paper we will use the VXD formalism to interpret this  
 as a correspondence between XD dimensional geometry and quantum behavior. We will test such a description  with a very simple formalization of the Quark-Gluon-Plasma freeze-out.

The analogies between XD theory and QM are well known and were first noted by O. Klien when, in analogy with the Bohr-Sommerfeld quantization, he introduced the idea of cyclic XD to describe the quantization of the electric charge \cite{Klein:1926tv}.
We may think for instance of the analogies between the resolution of the mass eigensystem and energy eigensystem in:  a Kaluza-Klein (KK) field  and the quantization of a ``particle in a box'' \cite{Randall:1999vf};   a XD field
with  ``brane terms''  and   a  Schr\"odinger problem with  Dirac delta potentials \cite{Dvali:2001gm,Carena:2002me};
 an XD theory with ``soft-walls''  and  the semi-classical quantization of a harmonic oscillator \cite{Karch:2006pv} or in the Front-Light-Quantization \cite{Vary:2009qz,Honkanen:2010nt}. In forthcoming papers we will extend this correspondence  to interpret gauge symmetry breaking mechanisms typical of XD or strong interacting theories in terms of superconductivity, \cite{Dolce:2009cev4,Dolce:superc}, as well as  the behavior of electrons in carbon nanotubes and graphene in terms of \emph{virtual} KK modes . \footnote{Carbon nanotubes and graphene provide an exceptional arena  to probe our formalism of VXD. In graphene the electron behaves as 2D massless particles. As a direction is curled up as in nanotubes, even if the drift velocity along the axial direction is set to zero (rest frame), the electron can have a residual cyclic motion along the radial direction. As for the proper-time periodicity of de Broglie ``internal clock'' this defines the effective mass of the electrons in nanotubes \cite{Dolce:superc}.   Note however that the energy spectrum in carbon nanotubes 
 can not be directly derived from a simple KK theory with XD on a lattice of $N$ sites, as claimed in \cite{deWoul:2012ed}. This would give an energy spectrum in which the momentum $p_n$ of the $n$-th KK particles are unrelated  (non-\emph{collective} description of the KK modes).
  To obtain the  correct energy spectrum of carbon nanotubes in which the KK modes have a \emph{collective} behavior  it is necessary to assume that the compact XD is \emph{virtual}.  }

It is interesting to mention that, similarly to sec.(\ref{sec:Correspondence-with-QFT}), the correspondence to the quantum formalism can be extended to describe  ordinary KK theories in a Hilbert space.  The KK modes form a complete set with given inner-product and they can be described by the   Hilbert eigenvector $\left|\phi_{n}\right\rangle $. Thus we introduce a mass operator $\mathcal M$ whose its spectrum is the KK mass spectrum,  $\mathcal{M}\left|\phi_{n}\right\rangle _{}\equiv M_{n}\left|\phi_{n}\right\rangle$, the time evolution of a KK field along the XD is described by the Schr\"odinger equation $i\partial_{s}|\phi(s)\rangle=\mathcal{M} |\phi(s)\rangle$ or by the evolution operator $\mathcal{U}(s';s)=e^{{-{{i}}\mathcal{M} (s-s')}}$, intrinsic commutation relation $[s, \mathcal M] = i$, and so on. A similar notation is used in Light-Front-Quantization \cite{Vary:2009qz}.

In this section we want to combine together all the main correspondences of field theory in compact 4D discussed so far (and in other peer-review papers):

\begin{itemize}
\item[\emph{i})] the  correspondence between a field with persistent periodicity  and a free quantum system, discussed in sec.(\ref{sec:Correspondence-with-QFT}) for the FPI, see \cite{Dolce:2009ce,Dolce:tune}; 
\item[\emph{ii})] the dualism between a field with persistent periodicity  and a field in flat XD,  obtained in  sec.(\ref{sec:Virtual-Extra-Dimension})  through the assumption of VXD, \cite{Dolce:2009ce}; 
\item[\emph{iii})] the geometrodynamical approach to interaction  and the related modulation of periodicity of the field, described in  sec.(\ref{sec:Towards-Interactions}), \cite{Dolce:tune}; 
\item[\emph{iv})] the  holographic effective description of an XD theory and its analogy with a theory with VXD, summarized in (\ref{source:holo:act}), see \cite{Casalbuoni:2007xn}. 
\end{itemize}

The outcome of the combination of all these points will be a dualism between the holographic description of the classical configurations of a field in a deformed XD metric and the quantum behavior of the corresponding interaction scheme.

The  classical evolution  of a free periodic field, point \emph{i}),  is described by the FPI (\ref{periodic:path.integr:Oper:Fey}) with time independent Hamiltonian $\mathcal H$ --- here the results of sec.(\ref{sec:Correspondence-with-QFT}) are generalized to three spatial dimensions $\mathbf{x}$. In this free case the  integral $\int \mathcal{D} \mathbf{x}$ is trivial. In fact, the periodicity of an isolated system is persistent and the same Hilbert space, \emph{i.e.} the same inner product, is defined in every point of its evolution.  
In this case, by using (\ref{eq:elem:xt:evol:2}), (\ref{elemet:phase:space}) and (\ref{mark:element:PI}) (see
\cite{Dolce:2009ce} for more detail or the introduction to the correlation function given for example in \cite{Peskin:1995ev}) the generating functional $\mathcal{Z}$ 
(\ref{eq:Feynman:Path:Integral}) can be formally written as 
\begin{equation}
\mathcal{Z} 
= \sum_{n}e^{-i p_{n}\cdot (x_{f}-x_{i})} 
=  \sum_{n}e^{-i M_{n} (s_{f}-s_{i})}
\leftrightarrow e^{i{\mathcal{S}}_{cl}(t_{f},t_{i})}\,.\label{flatVXD:freeQFT}
\end{equation}
This summarized the correspondence \emph{i}) between a field with persistent periodicity,including all its harmonics, and the quantum evolution of the corresponding free elementary system, sec.(\ref{sec:Correspondence-with-QFT}). 

In addition to this, we have the dualism, point \emph{ii}), between  a periodic field solution with persistent periodicity and  the classical configuration of a field in  flat XD. This dualism, investigated in sec.(\ref{sec:Virtual-Extra-Dimension}), is manifest if we assume that the XD is \emph{virtual}, \ie if we impose (\ref{proptime:realtime}). It  can be summarized as \begin{equation}
{\mathcal{S}}_{cl}(t_{f},t_{i})\leftrightarrow {\mathcal{S}}_{cl}^{5D}(s_{f},s_{i})\,.\label{VXD:5D:dualism}
\end{equation}
As shown in (\ref{flatVXD:freeQFT}), this can also be seen  if we evaluate the phase of the field in the rest frame $p_{n}\cdot (x_{f}-x_{i}) =  M_{n} (s_{f}-s_{i})$ in which the 4-momentum spectrum leads to a mass tower, see (\ref{eq:4D:debroglie:period}) and (\ref{eq:normal:ordered:spectr}).

Hence, the combination of point \emph{i}) and point \emph{ii}), \ie (\ref{flatVXD:freeQFT}-\ref{VXD:5D:dualism}), can be expressed as a  correspondence between classical configurations of a field in  flat VXD and ordinary quantum behavior of an isolated elementary system 
 \begin{equation}
\mathcal{Z}=\int_{V_{\mathbf{x}}} {\mathcal{D} \mathbf{x} }e^{i  \mathcal{S}_{cl}(t_{f},t_{i})}  \leftrightarrow e^{i {\mathcal{S}_{cl}^{5D}}(s_{f},s_{i})} \,.\label{eq:FPI-free:XD:flat}\end{equation}
According to our theory the case of an infinite VXD describes the quantum behavior of a massless system ($M_n = 0$), whereas the case of a compact VXD describes the quantum behavior of a massive  system ($M_n \neq 0$). Therefore, similarly to AdS/CFT, a compact VXD (finite periodicity) explicitly breaks the conformal invariance of the corresponding QFT describing  free elementary massive system.

Now we consider the geometrodynamical description  of interactions, point \emph{iii}), given in sec.(\ref{sec:Towards-Interactions}), \emph{i.e} modulation of periodicity,  in order to generalize (\ref{eq:FPI-free:XD:flat}). The combination of \emph{i}), \emph{ii}), and \emph{iii}) will result in a   more general correspondence between the classical configurations of a field  in deformed-VXD
and the quantum behavior  of the corresponding interaction scheme.
  
To see this we must formalize the propagation of a periodic field with locally modulated 4-periodicity $T'^\mu(x)$. As shown in \cite{Dolce:tune}, its resulting evolution is described by the ordinary FPI of the corresponding interaction scheme. 
The derivation of the  FPI (\ref{eq:Feynman:Path:Integral}) can in fact be  generalized to  the classical evolution of a modulated de Broglie ``periodic phenomenon'' as follows.

For a periodic field with persistent periodicity we can define a homogeneous 4-momentum operator $\mathcal{P}_\mu = \{\mathcal H, - \mathcal{P}_{i}\}$. Similarly to (\ref{eq:hamilt:def:gen:state}-\ref{eq:moment:def:gen:state}), the locally modulated spectrum of an interacting periodic field can be described by a corresponding non-homogeneous 4-momentum operator $\mathcal{P}'_{\mu}(x)$. 
As shown in detail in \cite{Dolce:tune}, the local operator $\mathcal{P}'_{\mu}(x)$ can be obtained from the homogeneous $\mathcal{P}_{\mu}$ through the transformation of 4-momentum associated with the interaction scheme (\ref{eq:deform:4mom:generic:int}):  
$
\mathcal{P}_{\mu} \rightarrow \mathcal{P}'_{\mu}(x) = e_{\mu}^{a}(x)  \mathcal{P}_{a} 
$, \cite{Dolce:tune}. 

Moreover,  the space-time evolution of a locally modulated periodic field solution is still Markovian (unitary), as shown in \cite{Dolce:tune}. That is, the total 4D evolution can be described as the product of elementary 4D evolutions, and the elementary 4D evolutions of an interacting cyclic field are formally described by (\ref{elemet:phase:space}), provided that the homogeneous operators are substituted by the corresponding local ones: $\mathcal H \rightarrow \mathcal H'$  and  $\mathcal P_{i} \rightarrow \mathcal P'_{i}$.
 Furthermore, we must consider that in case of interaction  the spatial periodicity is not persistent but it is modulated from point to point. Thus, to each local elementary 4D  evolution is associated a different local Hilbert space. In fact, the inner product (\ref{Hilbert:innerprod}) must be locally substituted by $\int^{V'_\mathbf{x}(\mathbf{X})} \frac{d \mathbf{x}'(\mathbf{X})}{V'_\mathbf{x}(\mathbf{X})}$. 
Nevertheless, a relativistic interaction can be supposed as limited to a (finite or infinite) region of space $\mathcal{I}$. If the integration volume $V'_\mathbf{x}(\mathbf{X})$ is bigger (or infinite) than the interaction region $\mathcal{I}$, \emph{i.e.} if we integrate over a large or infinite number of periods $N_{T}$,  we can  assume that the volume $V'_\mathbf{x}(\mathbf{X})$, as well as the normalization of the fields, is overall not affected by these local deformations. That is, for a sufficiently large integration region we have $V'_\mathbf{x}(\mathbf{X}) \cong V_{\mathbf{x}}$. As already shown in \cite{Dolce:tune}, we find out that the correct mathematical tool to describe these locally modulated elementary evolutions in which the inner product varies from point to point is again the integral $\int \mathcal{D} \mathbf{x}$ (which in case of interactions is not trivial).   

For a consistent generalization of (\ref{eq:Feynman:Path:Integral}) to interaction we must consider  that the action $ \mathcal{S}_{cl}(t_{f},t_{i})$, describing a classical isolated particle, is locally transformed under interactions.  
Similarly to (\ref{class:pt:action}) and (\ref{class:pt:lagrang}) we find that the non-homogeneous Hamiltonian  and momentum operator  $ \mathcal H'$  and  $ \mathcal P'$  define the Lagrangian ${L}'_{cl}\equiv\mathcal{P}'_{i}  \dot \mathrm{x}^{i}_{m } - \mathcal{H}'$, and in turn the transformed action $\mathcal{S}'_{cl}(t_{b},t_{a})\equiv\int_{t_{a}}^{t_{b}}dt {L}'_{cl}$. 
As already said,  $\mathcal{P}'_{\mu}(x)$ transforms 
 as the 4-momentum $\bar p'_{\mu}(x)$ of the corresponding classical particle (\ref{eq:deform:4mom:generic:int}). Thus, we find that 
  the transformed action  $\mathcal{S}'_{cl}(t_{b},t_{a})$ is formally the action of the corresponding interacting classical particle (written in terms of operators).

The demonstration used to derive  (\ref{eq:Feynman:Path:Integral}) can now be generalized to interaction. In fact,\cite{Dolce:tune}, we can plug point by point the local completeness relations of the energy eigenmodes in between the elementary 4D evolutions. Since in case of interaction the elementary 4D evolutions are described by the local operators $ \mathcal H'$  and  $ \mathcal P'$,  the FPI (\ref{eq:Feynman:Path:Integral}) turns out to be generalized to  
 \begin{equation}
\mathcal{Z}=\int_{V_{\mathbf{x}}} {\mathcal{D} \mathbf{x} }e^{ i  \mathcal{S}'_{cl}(t_{f},t_{i})}\,.\label{eq:Feynman:Path:Integral:Inter}
\end{equation}
Indeed this is the ordinary FPI associated to the interaction scheme (\ref{eq:deform:4mom:generic:int}).
In the l.h.s. of the correspondence (\ref{eq:FPI-free:XD:flat}), 
 interaction implies the following formal substitution of action: 
$\mathcal{S}(t_{f},t_{i}) \rightarrow  \mathcal{S}'(t_{f},t_{i})$.

Similarly we want to generalize to interaction the {r.h.s.} of the correspondence (\ref{eq:FPI-free:XD:flat}). According to sec.(\ref{Geometrodynamics:boundary}),  the
evolution of an interacting periodic field is  
 also dual to the classical configurations of a field
in the corresponding deformed XD metric (\ref{VXD:deform:metric}). The dualism is manifest if we assume that the XD is \emph{virtual}. Therefore,  in the {r.h.s.} of  (\ref{eq:FPI-free:XD:flat}), the interaction can be represented  by the  formal substitution of the  action in flat XD with the action in corresponding deformed XD,  
$\mathcal{S}^{5D}_{cl}(s_{f},s_{i}) \rightarrow \mathcal{S}'^{5D}_{cl}(s_{f},s_{i})$.  
Hence, combining \emph{i}), \emph{ii}), and \emph{iii}), we have found that the classical configuration of a bosonic field in a particular deformed VXD background describes the quantum behavior of the corresponding interaction scheme. 

Now we can apply the holographic prescription, point \emph{iv}), to describe  the 5D theory $\mathcal{S}'^{5D}_{cl}(s_{f},s_{i})$ in an effective way. By applying the holographic prescription to a 5D theory, whose XD  is interpreted as \emph{virtual}, we will get an effective description of the quantum behavior of the corresponding interacting system. 
That is, similarly to the boundary action in VXD (\ref{eq:act:virt:hologr}), the holographic prescription in VXD consists in  projecting the bulk dynamics of the action (\ref{5D:massless})  on the boundary. 
 As in ordinary holography the source field can be regarded as the fundamental mode of the  tower. Thus the holographic formulation turns out to be an effective description of the VXD theory at energies
$E^{eff}$ smaller than or of the order  of the fundamental
energy $\bar{E}$ (this is similar to the projection of a Euclidean partition function on the ground state at large $\beta$ \cite{Zinn-Justin:2000dr}). That is, \cite{Gherghetta:2006ha,Contino:2004vy,Barbieri:2003pr,Casalbuoni:2007xn},
\begin{equation}
{\mathcal{S}}_{cl}^{5D}(s_{f},s_{i})\sim\mathcal{S}^{Holo}_{\Phi|_{\Sigma}=e\phi_{\Sigma}}(s_{f},s_{i})+\mathcal{O}(E^{eff}/\bar{M})\,.\label{VXD:holo:appr}
\end{equation} 
 In an ordinary XD theory, holography  provides an effective and  \emph{collective} description of the propagation of the KK modes. However, as already noted  at the end of sec.(\ref{sec:AdSCFT:interpr}), by assuming a VXD, such a \emph{collective} description is already explicit, even without holography. In fact the \emph{virtual} KK modes naturally describe  the quantum excitations of the same fundamental system (string), \ie they are not independent fields. On the other hand, the fundamental level of the \emph{virtual} KK tower represents the non-quantum  limit of the field in VXD (\emph{i.e.} it can be locally matched with a classical KG mode).  Therefore, if we explicitly assume a VXD, the only effect of the holographic prescription is an effective description of the corresponding quantum behavior. 
The holographic action $\mathcal{S}^{Holo}_{\Phi|_{\Sigma}=e\phi_{\Sigma}}(s_{f},s_{i})$ of (\ref{source:holo:act}), including the kinetic term which describes the ``dynamics'' of the source field $\bar \Phi|_{\Sigma} \sim e \phi_{\Sigma} $, can be thought of as the effective action $\mathcal{S}^{eff}_{\Phi|_{\Sigma}} $ of ordinary perturbation theory: $\mathcal{S}^{Holo}_{\Phi|_{\Sigma}} \sim \mathcal{S}^{eff}_{\Phi|_{\Sigma}} $. This actually corresponds to a first order expansion in $\hbar$ and avoids  the use of the functional integral in the generalization to interaction of the {r.h.s} of the correspondence (\ref{eq:FPI-free:XD:flat}) (the dependence on $s$ has been eliminated). 
 
Finally  we can combine points \emph{i}), \emph{ii}), \emph{iii}), and \emph{iv}), obtaining an effective dualism between the holographic description of the classical configurations of a field in a deformed XD background --- with infinite  or compact XD --- and the quantum behavior of an interacting quantum system ---  massless (conformal) or massive (non-conformal) respectively. This dualism is manifest if we assume that the XD is \emph{virtual}. Therefore, the generalization of both sides of the correspondence (\ref{eq:FPI-free:XD:flat})  to interaction can be 
 formally summarized in terms of the holographic prescription \footnote{In the action $\mathcal{S}'$ on the {l.h.s.} of this correspondence a source term is understood. In fact, in (\ref{VXD:holo:appr}), through the source field $\phi_{\Sigma}$, we are explicitly selecting a particular interacting cyclic field solution with fixed coefficient $a_{n}$ and fundamental mode $\bar \Phi$.} by the following  mnemonic correspondence  
\begin{equation}
\mathcal{Z}=\int_{V_{\mathbf{x}}} {\mathcal{D}\mathbf{x}} e^{ i\mathcal{S}'(s,s')} \leftrightsquigarrow 
e^{ i \mathcal{S}^{Holo}_{\Phi|_{\Sigma}=e\phi_{\Sigma}}(s,s')}\,.\label{VXD:QFT:corr}\end{equation}

The resulting dualism from \emph{i}), \emph{ii}), \emph{iii}), and \emph{iv}) is therefore reminiscent of the phenomenological formulation of the AdS/CFT correspondence (\ref{AdS:CFT:corr}). 
This unconventional description of AdS/CFT will be tested in the next section for the specific case of a simple model of QGP freeze-out obtaining analogies with some basic aspects of AdS/QCD.\footnote{In  \cite{Dolce:tune} we show how  to retrieve gauge
  interactions from particular geometrodynamics of the compact 4D, in what can be regarded as a formulation  of Kaluza's original proposal \cite{Kaluza:1921tu} in VXD. In this case the correspondence between classical geometry and quantum behavior, (\ref {periodic:path.integr:Oper:Fey})  will formally reproduce the ordinary FPI of scalar QED.} 
  
In this paper we will limit the application of this correspondence to the study  of warped XD geometry. However,  the same approach has been used \cite{Dolce:tune} to reveal the geometrodynamics scalar associated to scalar QED. Remarkably, the interpretation of this geometrodynamical description of gauge interaction in terms of the VXD formalism leads to interesting analogies with Kaluza's original proposal \cite{Kaluza:1921tu}. That is, in case of gauge interaction the corresponding VXD metric is a Kaluza-like metric.

\subsection{Quark-Gluon-Plasma freeze-out}

A possible phenomenological application of  the  classical to quantum correspondence discussed above
is represented by  the simple case of the QGP exponential freeze-out as described (at a classical level) by the Bjorken Hydrodynamic
Model (BHM)  \cite{Magas:2003yp} or by thermal QCD \cite{Satz:2008kb}.  We therefore consider a collider experiment
in which we imagine having a volume of quarks and gluons originally
at high temperature, \emph{i.e.} at high energy $\bar{E}^{UV}=\Lambda$
and high spatial momentum $|\bar{\mathbf{p}}^{UV}|$. We assume that the  
QGP is in first approximation composed by massless
fields, so that, (\ref{eq:normal:ordered:spectr}), the energy and the momentum vary \emph{conformally} during the freeze-out
$\bar{E}^{UV}\sim |\bar{\mathbf{p}}^{UV}|$. 

During the freeze-out, the system radiates
energy hadronically or electromagnetically as long as the temperature
of the fields is higher than that of the surrounding environment \cite{Satz:2008kb}. 
In terms of de Broglie 4-periodicity (\ref{eq:4D:debroglie:period}),
 this  means that the fields inside the QGP  initially have
small time periodicity $T_{t}^{UV}$ and spatial periodicity $\vec \lambda_{x}^{UV}$. These are modulated \emph{conformally} during the freeze-out, \ie $T_{t}\sim| \vec \lambda_{x}|$, whereas the world-line parameter can be regarded as non-compact (infinite periodicity).

According to the simple BHM model  \cite{Magas:2003yp}, during the freeze-out 
the energy density $\eps(s)$  has an exponential gradient with respect to the laboratory time, \ie proper-time, $s$. That is, the energy of the fields decays exponentially during the freeze-out $\bar E \rightarrow \bar{E}(s) = e^{-ks}\bar{E}$. Thus $k$ describes the gradient of the freeze-out.
It is interesting to note that in  the analogy between  QGP and a thermodynamic system \cite{Satz:2008kb}, this corresponds to an exponential decay of the temperature $\mathcal T (s)$ of the fields which can also be interpreted in terms of Newton law's of cooling:   $\partial_{s}\mathcal T (s) = - \mathcal T (s)/ k$ with gradient $k$. 
Therefore the thermodynamic system passes  from a
high temperature state characterized by small time periodicity
$T{}_{t}^{UV}$, to a  state characterized by large
 periodicities $T_{t}^{IR}$ ($T{}_{t}^{UV}\ll T_{t}^{IR}$). The possible application of our field theory with intrinsic periodic periodicity to describe a thermodynamical system has been discussed in \cite{Dolce:2009ce,Dolce:superc} and will be extended in future papers. We will not give the detail about this, but it can be easily figured out if we consider that the Minkowskian periodicity and the corresponding quantized energy spectrum of our theory play a role analogous to the ``mathematical trick'' of the Euclidean time and the Matsubara frequencies of finite temperature field theory, respectively.

In the approximation of massless fields, this means that, according to the simple BHM model  \cite{Magas:2003yp} prescribing in the QGP freeze-out,  the 4-momentum of the fields decreases conformally and exponentially  
\begin{equation}
\bar{p}_{\mu}\rightarrow\bar{p}'_{\mu}(s)\simeq e^{-ks}\bar{p}_{\mu}\,.\label{eq:deform:4mom:QGP}
\end{equation}

Indeed the interaction scheme associated with the QGP freeze-out is described by a conformal exponential of the  world-line parameter
$s$ and the  parameter $k$. In terms of our geometrodynamical description of interaction   (\ref{eq:deform:4mom:generic:int}), the freeze-out is  therefore encoded  by the conformal tetrad 
\begin{equation}
e^{a}_{\mu}(s)\simeq \delta^{a}_{\mu} e^{-ks}\,.\label{eq:deform:tetrad:QGP}
\end{equation}

Proceeding in analogy with the description of interaction given in par.(\ref{sec:Towards-Interactions}),
from (\ref{eq:deform:4mom:QGP}) and (\ref{eq:deform:4period:generic:int}) we find that during the freeze-out the 4-periodicity has an exponential and conformal
dilatation
\begin{equation}
T^{\mu}\rightarrow T'^{\mu}(s)\simeq e^{ks}T{}^{\mu}\,.\label{eq:deform:4period:QGP}
\end{equation}
According to (\ref{eq:deform:mesure:generic:int}), this modulation of periodicity is encoded in the substitution of variables
\begin{equation}
dx_{\mu}\rightarrow dx'_{\mu}(s)\simeq e^{-ks}dx{}_{\mu}\,.
\end{equation}
Thus, (\ref{eq:deform:metric:generic:int}), the QGP freeze-out is encoded by the warped metric $ds^{2}=e^{-2ks}dx{}_{\mu}dx{}^{\mu}$.

As inferred in (\ref{VXD:deform:metric}), by treating the world-line parameter $s$  as a VXD, the exponential dilatation of the 
4-periodicity during the QGP freeze-out of massless fields  ($dS^{2}\equiv0$) can be equivalently encoded in the
\emph{virtual} AdS metric\begin{equation}
dS^{2}\simeq e^{-2ks}dx{}_{\mu}dx{}^{\mu}-ds^{2}\equiv0~.
\end{equation}

It is interesting to note that the de Broglie-Planck relation  (\ref{eq:4D:debroglie:period})  explicitly implies that the 
exponential evolution of the energy $\bar{E}(s)$ of the fields from
$s^{UV}$ to $s^{IR}$ is  parameterized, through the Planck constant, by the inverse of the time periodicity
$T_{t}(s)$. That is, from  (\ref{eq:deform:4mom:QGP}-\ref{eq:deform:4period:QGP}) we find that
 the characteristic time periodicity $T_{t}(s)$  of the QGP during the freeze-out  is nothing other than the  conformal parameter of AdS/CFT
\begin{equation}
z(s)\equiv\frac{e^{ks}}{k}=T_{t}(s)=\frac{1}{\bar E(s)}\,.\end{equation}
In agreement with the AdS/CFT dictionary, we see that  $z$ naturally parameterizes the inverse of the energy of the QGP fields because it actually corresponds to the de Broglie time periodicity of the fields.  

We use $z^{UV}$ and $z^{IR}$ to denote a UV
energy scale $\Lambda$ (Planck scale) and  an IR energy scale $\mu$
(TeV scale), respectively, \begin{equation}
T_{t}^{UV}=\frac{1}{\Lambda}=\frac{e^{ks^{UV}}}{k}\,\,\,,\,\,\,\,\, T_{t}^{IR}=\frac{1}{\mu}=\frac{e^{ks^{IR}}}{k}\,.\end{equation}
According to our geometrodynamical description, as well as to the Bjorken and thermal QCD description, 
the \emph{virtual} AdS background leads to an interaction scheme in which the 4-periodicity of the QGP particles has exponential dilatation, \ie  the 4-momentum decays exponentially. 
In  this  approximative description of interaction  in which the masses are neglected, the kinematical information
is encode on the VXD metric, in particular by the AdS curvature.
The quantum evolution
of the strong coupling constant can be qualitatively obtained in a way that resembles AdS/QCD. As already said in par.(\ref{sub:VXD/QFT--correspondence}) and summarized by (\ref{VXD:QFT:corr}), from
the resulting correspondence between the usual QFT and compact 4D field
theory investigated in sec.(\ref{sec:Correspondence-with-QFT}) and
  sec.(\ref{sec:Virtual-Extra-Dimension}), we in fact expect to find out that the exponential
dilatation of the 4-periodicities written in terms of the
conformal parameter $z$ describes the quantum behavior of the QGP
at different energy scales. 

This correspondence indeed resembles  AdS/QCD. The topic is too wide to be treated  in detail here.
However, our strategy to describe the quantum
behavior arising from the classical geometrodynamics associated to (\ref{eq:deform:4period:QGP}) will be to follow the line of some  of the founding papers on the AdS/QCD correspondence \cite{ArkaniHamed:2000ds,Pomarol:2000hp,Son:2003et,DaRold:2005zs}.
 In this way we will find out an unconventional interpretation of basic aspects of the correspondence similarly to the interpretation already given for the conformal parameter $z$ and the curvature $k$. 
 According to our theory, the quantum behavior of massless fields  ($M_n = 0$) is described by an infinite warped VXD (infinite proper-time periodicity of the fields). This means that the conformal parameter $z$, \emph{i.e.} the time periodicity of the fields  is allowed to take any value.   Since in this case we are not  introducing any scale, the energy of the  fields in VXD is allowed to vary from zero to infinity.
 
  As shown in sec.(\ref{sub:VXD/QFT--correspondence}), a non-compact XD geometry describes the quantum behavior of conformal fields\footnote{Since the mass has a geometrical meaning in the de Broglie ``periodic phenomenon'', it can be shown that the terminology ``conformal invariance''   has an extended validity for cyclic fields. We will give more detail about this in future publications}.  
 Here we consider the case of  a 5D gauge theory with bulk coupling $g_{5}$ and  an infinite VXD. In this way all the modes  are flat, \emph{i.e.} independent  on the VXD $A^{(flat)}_{ \mu}(x,z)=A^{(flat)}_{ \mu}(x)$. At this point,  similarly to (\ref{eq:4d:periodic:field}), the quantum behavior associated to the classical configurations  of the fields in this \emph{virtual} AdS metric can be extracted by integrating out the VXD  (i.e. the conformal parameter describing the periodicity of the QGP), so that,  for every (massless) mode of the tower   we get 
\begin{eqnarray}
\mathcal{S}^{AdS} & \simeq & -\int_{1/\Lambda}^{1/\mu}\frac{dz}{kz}\int \frac{d^{4}x}{4g_{5}^{2}} F^{(flat)}_{\mu\nu}(x,z)F^{(flat) \mu\nu}(x,z)\nn \\ 
 & = & -\frac{\log\frac{\mu}{\Lambda}}{4kg_{5}^{2}}\int d^{4}xF^{(flat)}_{\mu\nu}(x)F^{(flat) \mu\nu}(x)\,.
 \end{eqnarray}
Hence, \cite{ArkaniHamed:2000ds,Pomarol:2000hp}, the effective 4D
coupling behaves logarithmically with respect to the infrared scale
\begin{equation}
g^{2}\simeq\frac{g_{5}^{2}k}{\log\frac{\mu}{\Lambda}}\,.\label{log:run:ZMA}\end{equation}
This reproduces the quantum behavior with the energy
of the strong coupling constant, \aka the \emph{asymptotic freedom}, as long as we suppose \cite{Son:2003et,Pomarol:2000hp,DaRold:2005zs}
\begin{equation}
\frac{{1}}{{k}}\sim\frac{{N_{c}g_{5}^{2}}}{{12\pi^{2}}}\,.\label{matching:curvature:AdSQCD}
\end{equation}
Thus,  according to our description, the curvature encodes the strength of  the interaction, \emph{i.e.} the gradient of the QGP freeze-out.
In agreement with the AdS/CFT dictionary, the case of infinite VXD actually corresponds to  a conformally invariant quantum theory.

The analogy with AdS/CFT phenomenology can also be extended to the case of compact VXD. In an Orbifold description, the PBCs  can be substituted\footnote{Since in the Hilbert space (\ref {Hilbert:innerprod}) of our theory of
  topology $\protect \mathbb {S}^{1}$ only the modulo of the field has physical
  meaning, we can adopt the formalism of XD field theory in an Orbifold
  $\protect \mathbb {S}^{1}/\protect \mathcal {Z}_{2}$ and describe the PBCs in
  terms of N-N BCs (or D-D BCs depending if we want to consider or not the
  purely translational mode $n=0$).} with N and N BCs at the IR and UV energy scales, respectively.\footnote{By assuming explicit branes the QGP freeze-out is limited between an initial and final
  temperature (the branes could be interpreted
  as phase transition points).}  According to our interpretation, the \emph{virtual} KK modes of this VXD theory are the quantum excitations of the same fundamental conformal system, in fact they can be regarded as describing the hadrons as we will see below.  
However, according to our theory, a compact VXD  with PBCs generates a mass spectrum, which is non-trivial in the case of warped VXD \footnote{The massless mode of this model behaves as in the conformal case above \cite
  {ArkaniHamed:2000ds}. This Zero Mode Approximation can be obtained by
  substituting the NBCs with appropriate SBCs in order to eliminate all the
  higher massive modes.}.
 This  means that the conformal invariance is broken in the corresponding quantum theory, in agreement with the  AdS/CFT dictionary. 
The quantum behavior can be described in an effective way by applying the holographic prescription to this compact VXD geometry. In fact, as already said, holography implicitly provides  an effective description of the  propagation of the  \textit{virtual} KK modes. That is, the KK modes are already described in an effective \emph{collective} way in the holographic propagator.   
 Thus, we introduce an IR source 
 field $\bar A_{\mu}(q)$ \cite{ArkaniHamed:2000ds} with coupling $e$,
 whereas we keep NBCs at the UV scale.  
As argued 
in par.(\ref{sub:VXD/QFT--correspondence}) the assumption of a source field at the IR brane  corresponds to an effective description of  NBCs  \cite{Son:2003et}, \emph{i.e.} of the underlying theory with compact VXD and N-N BCs.  
Hence, by assuming  Euclidean momentum ($q\rightarrow iq$) and expanding in the
limit of large 4-momentum with respect to the IR scale ($\Lambda\gg|q|\gg\mu$), we find out that
the leading behavior of the holographic correlator of the action  (\ref{VXD:holo:appr})  is \begin{equation}
\Pi^{Holo}(q^{2})\sim-\frac{q^{2}}{2kg_{5}^{2}}\log\frac{q^{2}}{\Lambda^{2}}~.\end{equation} 
By identifying the source field $\bar A_{\mu}$ with the vectorial field of a chiral theory,
$\Pi^{Holo}(q^{2})$  
approximately matches
the vector-vector two-point function of ordinary QCD. 
Analogously to (\ref{log:run:ZMA}) this gives an estimation of the logarithmic running of the gauge coupling
of  QCD, \ie asymptotic freedom,  
\begin{equation}
\frac{1}{e_{eff}^{2}(q)}\simeq\frac{1}{e^{2}}-\frac{{N_{c}}}{{12\pi^{2}}}\log\frac{q}{\Lambda}\label{log:run:holo}~.
\end{equation}
This quantum behavior has been obtained without imposing  any
explicit quantization except BCs.

We must note that, even though we have generated massive modes by imposing a compact VXD, we are still adopting a  conformal metric in which  $dt^{2} \sim d x^{2}$. But according to (\ref{relat:constr:4T}) and  (\ref{VXD:deform:metric}),  this conformal behavior is justified only in the massless approximation. That is, the AdS metric is not fully consistent with a massive dispersion relation since it describes a conformal deformation of the space-time dimensions. This reveals  the limit of this formulation of the QGP and at the same time  it allows us to speculate about the fact that  deformations of the AdS metric, for instance through the assumption of a dilaton in (\ref{VXD:deform:metric}) or ``soft-walls'',  provide more realistic predictions of the hadronic mass spectrum 
 \footnote{In a bottom-up approach to AdS/QCD, one can implement 5D chiral models with a
  soft-wall in the IR or with a dilaton which give realistic predictions for
  the masses and for the decay constants of light vector and axial-vector
  mesons, for their form factors, for the vector meson dominance, for the
  Weinberg sum rules, etc, in sufficient agreement with the experimental data
  \cite
  {Pomarol:2000hp,ArkaniHamed:2000ds,Son:2003et,Karch:2006pv,Erlich:2005qh}.}. In future studies we will try to interpret the meaning of these deformations and the
  related Regge-like behavior of the hadronic mass spectrum in terms of
  Kaluza's matrix and winding numbers around compact 4D. 
As further evidence of the relevance of semi-classical methods, based on PBCs, in  AdS/QCD phenomenology, we mention that  the ``soft-wall'' was originally justified \cite{Karch:2006pv} in terms the  semi-classical  Bohr-Sommerfeld quantization---\emph{i.e.} a periodicity condition---associated with a harmonic potentials.  Similar results are obtained in Light-Front quantization in which PBCs at the ends of the Light-Front coordinate are imposed as semi-classical quantization condition, \cite{Vary:2009qz}.

The semi-classical quantization by PBCs of field theory in compact 4D and its (unconventional) interpretation in terms of AdS/CFT presented in this paper must be tested for other known aspects of
the AdS/QCD. Nevertheless the remarkable correspondences between 5D geometry and 4D quantum behavior pointed out in this paper can provide an intuitive tool to develop more  realistic phenomenological AdS/QCD models. In future papers we will extend this analysis by using the analogy (\ie mode expansion associated to BCs) with the Light-Front quantization and ``soft-walls'' (see below).

Indeed we have a description of the AdS/CFT which, on the one hand, mimics the dilatation of periodicity of reference clocks  and red-shift typical of a gravitational interaction, as described in  par.(\ref{relativistic:geometrodynamics}) and on the other hand, it can be interpreted in terms of wave-particle duality, through   the  ``periodic phenomenon''. 

\subsection{Comments and  Outlooks}

Finally we introduce another intuitive element to interpret the good properties and the consistency of field theory in compact 4D  and fundamental topology $\mathbb S^1$. By imposing the constraint of intrinsic periodicity, the resulting field solutions $\Phi(x)$ of the theory can be regarded as a relativistic string vibrating  in compact 4D. The  harmonics resulting from its space-time vibrations, together with the corresponding  energy-momentum spectrum, can be regarded as quantum excitations according to the results of sec.(\ref{sec:Correspondence-with-QFT}). Remarkably this picture turns out to be the full relativistic generalization of sound theory. A sound signal is a packet of classical waves whose harmonic spectrum is determined by the vibration on compact spatial dimensions of a source, for instance a one-dimensional string. Loosely speaking, in our formalization of the de Broglie ``periodic phenomenon'' every matter particle can be regarded as a relativistic string (\ie governed by relativistic wave equations) vibrating  in compact 4D (in particular, with PBCs). These generate (unconstrained) relativistic waves playing the role of  the (massless) mediators of the interaction and thus defining the underlying reference space-time for a relational description of the sources (\eg the atomic orbitals can be regarded as the harmonics  of space-time vibrations of topology $\mathbb S^1 \times \mathbb S^2$). It is also interesting to point out that, from a historical point of view, the formalism of QM has its origin in the formalism of sound theory developed by Rayleigh at the end of the 19th century.

The idea to describe nature in terms of the mathematics of vibrating strings has indeed solid historical origins (which can be postdated to Pythagoras). This idea is also at the origin of modern String Theory (ST) whose mathematical beauty has attracted the interest of several generations of physicists. 
It is interesting to mention that, our formalism being  based on the assumption of a compact  world-line parameter, we have a fundamental analogy with  ST. In fact, ST is based on the fundamental assumption  that one of the two world-sheet parameters is compact (with PBCs, or  D-BCs and N-BCs, in the case of  closed strings or   open strings, respectively). In our theory, the compact world-line parameter, whose length is determined by the mass, plays a role similar to the compact world-sheet parameter of ST, so that  the bosonic action in compact 4D defined in  (\ref{generic:actin:comp4D}) can be regarded as the prototype of a simple, purely 4D, ST.  
 In this way it is possible to figure out that our theory inherits some of the interesting mathematical properties of ST, without the problematic requirement of the actual existence of unobserved XDs in the theory.  A exhaustive description of the analogy with string theory is beyond the scope of this paper.  However, in \cite{Dolce:tune} we have already mentioned that the energy levels of a ``periodic phenomenon'' can be described in terms Virasoro algebra (with implicit commutation relations).
Furthermore, we may note that   the dualism to XD theories allows us to associate to our vibrating strings a Veneziano amplitude whose poles, for instance calculated through the holographic prescription,  depend on the  VXD geometry, \cite{Afonin:2011nw,Afonin:2011hk}. 
   In the holographic prescription  of a  warped VXD , assuming dilatons or soft-walls in order to reproduce Regge poles, we find that the mass spectrum and the propagation of \emph{virtual} KK modes approximately matches the  hadronic mass spectrum and  quantum two-point function of QCD, respectively, \cite{Son:2003et}.
   Thus in our description the hadrons can be actually interpreted as  quantum excitations encoded in the harmonic modes of a vibrating string describing QCD. This interpretation of  QCD is at the very origin of string theory, but it is also well known that,  \cite{Chodos:1974je}, because of confinement, hadrons have a natural formalization in terms of field theory in compact 4D and BCs. 
 
 Our purely 4D reinterpretation of AdS/QCD can also be seen by considering the analogies  with Light-Front quantization. For instance, in the isotropy  QGP freeze-out should also involve the expansion in spherical harmonics associated to the  periodicities in $(\theta,\varphi)$.  Thus, the Light-Front coordinates, especially in the case of massless fields, represent a natural parameterization of the system, so that  our semi-classical quantization is achieved by imposing PBCs to the  Front-Light coordinates.  Indeed,  PBCs for the  Front-Light coordinates is at the base of  Light-Front quantization \cite{Vary:2009qz}.
  In this way, we can intuitively interpret the analogies between the Front-Light  quantization and  AdS/QCD, \cite{Vary:2009qz,Honkanen:2010nt}. As noticed recently,  Light-Front-quantization has also been  used to
  reproduce semi-classical results of perturbative QED \cite{Honkanen:2010nt} in terms of expansion of the harmonic modes allowed by the BCs.  In   agreement with the motivations of sec.(\ref{sec:AdSCFT:interpr}) and with the results of sec.(\ref{sec:Correspondence-with-QFT}), this can be regarded as an indirect confirmation of the other relevant results of field theory in compact 4D, that is to say the semi-classical description of scalar QED obtained in \cite{Dolce:tune}.  Another  relevant application of field theory in compact 4D would be a purely 4D reinterpretation of phenomenological XD models such as Randall-Sundrum, in terms of the formalism of VXD.  All these heuristic  considerations provide a general picture of the fundamental physical motivations for our formulation of field theory in compact 4D, and will be used in future research for other phenomenological applications  and predictions.

\addcontentsline{toc}{section}{Conclusions} 
\section*{Conclusions}

According to Witten, in the AdS/CFT correspondence, \emph{quantum behavior [...] are encoded in classical geometry}, \cite{Witten:1998zw}.
In this paper we have proposed an unconventional interpretation of fundamental aspects of this ``classical to quantum'' correspondence of AdS/CFT in terms of the undulatory mechanics and relativistic geometrodynamics, \cite{Dolce:2009cev4}.  
In recent papers, \cite{Dolce:2009ce,Dolce:tune}, we have formulated a possible realization of the so-called de Broglie ``periodic phenomenon'', at the base the wave-particle duality and whose character is ``yet to be determined'' \cite{Broglie:1924}. This formulation is based on the assumption of PBCs as semi-classical quantization conditions at the geometrodynamical boundary of a field theory defined in compact 4D.   The resulting semi-classical solution of the theory has remarkable formal correspondences with the fundamental aspects of ordinary QFT. It can also be regarded as a ``vibrating string'', and similarly to a  ``particle in a box'', its harmonics  describes the quantum excitation of the elementary system. 

  Field theory in compact 4D  has an exact dualism
to  XD field theory \cite{Dolce:2009ce}. In fact it can be formally retrieved  by identifying the cyclic XD of a 5D massless field with a  world-line parameter.   That is,  the cyclic XD of our purely 4D theory and its compactification length turns out to encode the so-called de Broglie ``internal clock''  and the Compton wavelength of a corresponding  particle, respectively. We address this identification by saying that field theory in compact 4D  has a 
\emph{virtual} XD. In this dual description  the KK modes are \emph{virtual} in the sense that they encode  quantum excitations.  
This idea is in the spirit of  Kaluza's and Klein's original proposals
\cite{Kaluza:1921tu,Klein:1926tv}:  Kaluza   introduced the XD as a ``mathematical trick'' rather than as a ``real'' XD whereas Klein used PBCs for a semi-classical interpretation of QM.

According to  undulatory mechanics, the variation of 4-momentum occurring during interaction can be equivalently described as modulations of 4-periodicity.    In the formalism of field theory in compact 4D, through the assumption of PBCs,  the modulation of 4-periodicity associated to a given interaction scheme can be described in terms of geometrodynamics of the boundary (similarly to the holographic principle), or equivalently in terms of deformations on a corresponding metric in \emph{virtual} XD. Moreover this geometrodynamical description of interaction mimics  linearized gravity. In fact, the spacial and temporal modulations of reference lengths and clocks are encoded in  local deformations of the space-time metric, respectively. Remarkably, this description typical of GR  can also be extended to describe gauge interactions in terms of local variations of reference frames, as proven in  \cite{Dolce:tune}.

Similarly to AdS/CFT, by combining the correspondence of field theory in compact 4D with ordinary QFT, its dualism to XD field theory, and the geometrodynamical description of interaction,  we have inferred that  he classical configuration of a field
 in a deformed  VXD reproduces the quantum behavior
of a corresponding interaction scheme. The application of this approach
to the exponential freeze-out of the QGP actually yields fundamental analogies with basic aspects AdS/QCD.

  \section*{Acknowledgements} 
 I would like to thank in chronological  order  V. Ahrens, M. Neubert, M. Reuter and T. Gherghetta   
 for fruitful and fair discussions,  and for their interest in new ideas in physics.   
  This paper is part of the project ``Compact Time and Determinism''. Its main results  have been 
partially published on-line in  \cite{Dolce:2009cev4}. 

\providecommand{\href}[2]{#2}\begingroup\raggedright\endgroup

\end{document}